\documentstyle[12pt]{article}
\begin{document}
\tolerance=5000
\def\be{\begin{equation}}
\def\ee{\end{equation}}
\def\bea{\begin{eqnarray}}
\def\eea{\end{eqnarray}}
\def\nn{\nonumber \\}
\def\cF{{\cal F}}
\def\det{{\rm det\,}}
\def\Tr{{\rm Tr\,}}
\def\e{{\rm e}}
\def\etal{{\it et al.}}
\def\erp2{{\rm e}^{2\rho}}
\def\erm2{{\rm e}^{-2\rho}}
\def\er4{{\rm e}^{4\rho}}
\def\etal{{\it et al.}}

\  \hfill
\begin{minipage}{3.5cm}
OCHA-PP-148 \\
NDA-FP-70 \\
\end{minipage}

\vfill

\begin{center}
{\large\bf Finite Action in d5 Gauged Supergravity and
Dilatonic Conformal Anomaly for Dual Quantum Field Theory}

\vfill

{\sc Shin'ichi NOJIRI}\footnote{nojiri@cc.nda.ac.jp},
{\sc Sergei D. ODINTSOV}$^{\spadesuit}$\footnote{
On leave from Tomsk State Pedagogical University, RUSSIA. \\
\ \hskip 1cm email: odintsov@ifug5.ugto.mx}, \\
{\sc Sachiko OGUSHI}$^{\heartsuit}$\footnote{
JSPS Research Fellow,
g9970503@edu.cc.ocha.ac.jp}
\\

\vfill

{\sl Department of Mathematics and Physics \\
National Defence Academy,
Hashirimizu Yokosuka 239, JAPAN}

\vfill

{\sl $\spadesuit$
Instituto de Fisica de la Universidad de 
Guanajuato \\
Apdo.Postal E-143, 37150 Leon, Gto., MEXICO}

\vfill

{\sl $\heartsuit$ Department of Physics,
Ochanomizu University \\
Otsuka, Bunkyou-ku Tokyo 112, JAPAN}

\vfill

{\bf ABSTRACT}

\end{center}

Gauged supergravity (SG) with single scalar (dilaton) and arbitrary scalar
potential is considered. Such dilatonic gravity describes special RG flows
in extended SG where scalars lie in one-dimensional submanifold of total
space. The surface counterterm and finite action for such gauged SG in
three-, four- and five-dimensional asymptotically AdS space are derived.
Using finite action and consistent gravitational stress tensor
(local surface counterterm prescription) the
regularized expressions for free energy, entropy and mass of d4 dilatonic
AdS black hole are found. The same calculation is done within standard
reference background subtraction.

The dilaton-dependent  conformal anomaly from d3 and d5 gauged SGs
is calculated using AdS/CFT correspondence.
Such anomaly should correspond to two- and four-dimensional dual
quantum field theory which is classically (not exactly)
conformally invariant, respectively.
The candidate c-functions
from d3 and d5 SGs are suggested. 
These c-functions which have fixed points in asymptoticaly 
AdS region are expressed in terms of dilatonic potential and 
they are positively defined and monotonic for number of potentials.
\newpage

\section{Introduction}

AdS/CFT correspondence \cite{AdS} may be realized in a sufficiently
simple form as d5 gauged supergravity/boundary gauge theory
correspondence.
The reason is very simple: different versions of five-dimensional
gauged SG (for example, $N=8$ gauged SG \cite{GRW} which contains
42 scalars and non-trivial scalar potential) could be obtained as
compactification (reduction) of ten-dimensional IIB SG. Then,
in practice it is enough to consider 5d gauged SG classical
solutions (say, AdS-like backgrounds) in AdS/CFT set-up
instead of the investigation of much more involved,
non-linear equations of IIB SG. Moreover, such solutions describe RG
flows in boundary gauge theory (for a very recent discussion of such
flows see \cite{CM,DF,NOR,lust,FGPW,GPPZ,BBHV} and refs. therein).
To simplify the situation in extended SG one can consider the
symmetric (special) RG flows where scalars lie in one-dimensional
submanifold of total space. Then, such theory is effectively
described as d5 dilatonic gravity with non-trivial dilatonic
potential. Nevertheless, it is still extremely difficult
to make the explicit identification of deformed SG solution with
the dual (non-conformal exactly) gauge theory.
As a rule \cite{DF,GPPZ}, only
indirect arguments may be suggested in such
identification\footnote{Such dual theory in massless case is,
of course, classically conformally invariant and it has
well-defined conformal anomaly. However, among the interacting
theories only ${\cal N}=4$ SYM is known to be exactly
conformally invariant. Its conformal anomaly is not renormalized.
For other, d4 QFTs there is breaking of conformal invariance
due to radiative corrections which give contribution
also to conformal anomaly. Hence, one can call such theories as
non-conformal ones or not exactly conformally invariant. The conformal 
anomalies for such theories are explicitly unknown. Only for 
few simple theories (like scalar QED or gauge theory without fermions) the 
calculation of radiative corrections to conformal anomaly has been done 
up to two or three loops. It is a challenge to find exact conformal anomaly.
 Presumbly, only SG description may help to resolve this problem.}.

 From another side, the fundamental holographic principle \cite{G}
in AdS/CFT form enriches the classical gravity itself (and here
also classical gauged SG). Indeed, instead of the standard
subtraction of reference background \cite{3,BY} in making
the gravitational action finite and the quasilocal stress tensor
well-defined one introduces more elegant, local surface
counterterm prescription \cite{BK2}. Within it one adds the
coordinate invariant functional of the intrinsic boundary
geometry to gravitational action.
Clearly, that does not modify the equations of motion. Moreover,
this procedure has nice interpretation in terms of dual QFT as
standard regularization. The specific choice of surface counterterm
cancels the divergences of bulk gravitational action. As a
by-product, it also defines the conformal anomaly of boundary QFT.

Local surface counterterm prescription has been successfully applied
to construction of finite action and quasilocal stress tensor on
asymptotically AdS space in Einstein
gravity \cite{BK2,myers,EJM,ACOTZ,Ben} and
in higher derivative gravity \cite{SNO}. Moreover,
the generalization to
asymptotically flat spaces is possible as it was first mentioned in
ref.\cite{KLS}.  Surface counterterm has been found for domain-wall
black holes in gauged SG in diverse dimensions \cite{CO}. However,
actually only the case of asymptotically constant dilaton
has been investigated there.

In the present paper we discuss the construction of finite action,
consistent gravitational stress tensor and dilaton-dependent Weyl
anomaly for boundary QFT (from bulk side) in three- and five-dimensional
gauged
supergravity with single scalar (dilaton)
on asymptotically AdS background. Note that dilaton is not
constant and the potential is chosen to be arbitrary.
The implications of results for
the study of RG flows in boundary QFT are presented, in particular,
the candidate c-function is suggested.

The next section is devoted to the evaluation of Weyl anomaly from
gauged supergravity with arbitrary dilatonic potential via AdS/CFT
correspondence. We present explicit result for d3 and d5 gauged SGs.
Such SG side conformal anomaly should correspond to dual QFT
with broken conformal invariance in
two and in four dimensions, respectively.
The explicit form of d4 conformal anomaly takes few pages,
so its lengthy dilaton-dependent
coefficients are listed in Appendix. The comparison with
similar AdS/CFT
calculation of conformal anomaly in the same theory but with constant
dilatonic potential is given. The candidates for c-function
in two and four dimensions are proposed.

Section three is devoted to presentation of acceptable proposal 
for candidate c-fnction given in terms of dilatonic potential. It is shown 
that for numberof potentials such c-function is monotonic and positively
defined.
It has fixed point in asymptotically AdS region. The comparison with
other c-functions is given.  

In section four we construct surface counterterms for d3 and d5 gauged SGs.
As a result, the gravitational action in asymptotically AdS space is finite.
On the same time, the gravitational stress tensor around such space is well
defined. It is interesting that conformal anomaly defined in second section
directly follows from the gravitational stress tensor with account of
surface terms.

Section five is devoted to the application of finite gravitational action
found in
previous section in the calculation of thermodynamical quantities in
dilatonic AdS black hole. The dilatonic AdS black hole is constructed
approximately, using the perturbations around constant dilaton AdS black hole.
The entropy, mass and free energy of such black hole are found using the
local surface counterterm prescription to regularize these quantities. The
comparison is done with the case when standard prescription: regularization
with reference
background is used. The explicit regularization dependence of the result is
mentioned. Finally, in the Discussion the summary of results is presented
and some open problems are mentioned.

\section{Weyl anomaly for gauged supergravity with
general dilaton potential}

In the present section the derivation of dilaton-dependent Weyl anomaly
from gauged SG will be given. As we note in section 4 this derivation can
be made also from
the definition of finite action in asymptotically AdS space.

We start from the bulk action of $d+1$-dimensional
dilatonic gravity with the potential $\Phi $
\be
\label{i}
S={1 \over 16\pi G}\int_{M_{d+1}} d^{d+1}x \sqrt{-\hat G}
\left\{ \hat R + X(\phi)(\hat\nabla\phi)^2
+ Y(\phi)\hat\Delta\phi
+ \Phi (\phi)+4 \lambda ^2 \right\} \ .
\ee
Here $M_{d+1}$ is  $d+1$ dimensional manifold whose
boundary is $d$ dimensional manifold $M_d$ and
we choose $\Phi(0)=0$. Such action corresponds to
(bosonic sector) of gauged SG with single scalar (special RG flow).
In other words, one considers RG flow in extended SG when scalars lie in
one-dimensional submanifold of complete scalars space.
Note also that classical vacuum stability restricts the form of
dilaton potential \cite{T}.
As well-known, we also need to add the surface terms \cite{3}
to the bulk action in order to have well-defined variational principle.
At the moment, for the purpose of calculation of Weyl anomaly (via
AdS/CFT correspondence) the surface terms are
irrelevant.
The equations of motion given by variation of (\ref{i}) with
respect to $\phi$ and $G^{\mu \nu}$ are
\bea
\label{ii}
0&=&-\sqrt{-\hat{G}}\Phi'(\phi)-\sqrt{-\hat{G}}V'(\phi)
{\hat G}^{\mu \nu}\partial_{\mu }\phi
\partial_{\nu}\phi \nn
&& +2 \partial_{\mu }\left(\sqrt{-\hat{G}}
\hat{G}^{\mu \nu}V(\phi)\partial_{\nu} \phi \right) \\
\label{iii}
0 &=& {1 \over d-1}\hat{G}_{\mu\nu}\left(
\Phi(\phi)+{d(d-1) \over l^2}\right)+\hat{R}_{\mu \nu}+V(\phi)
\partial_{\mu }\phi\partial_{\nu}\phi \ .
\eea
Here
\be
\label{iib}
V(\phi)\equiv X(\phi) - Y'(\phi)\ .
\ee

We choose the metric $\hat G_{\mu\nu}$ on $M_{d+1}$ and
the metric $\hat g_{\mu\nu}$ on $M_d$ in the following form
\be
\label{ib}
ds^2\equiv\hat G_{\mu\nu}dx^\mu dx^\nu
= {l^2 \over 4}\rho^{-2}d\rho d\rho + \sum_{i=1}^d
\hat g_{ij}dx^i dx^j \ ,\quad
\hat g_{ij}=\rho^{-1}g_{ij}\ .
\ee
Here $l$ is related with $\lambda^2$
by $4\lambda ^2 = d(d-1)/{l^{2}}$.
If $g_{ij}=\eta_{ij}$, the boundary of AdS lies at $\rho=0$.
We follow to method of calculation of  conformal anomaly
as it was done
in refs.\cite{NOano,NOOSY}
where dilatonic gravity with constant dilaton potential has been considered.
Part of results of this section concerning Weyl anomaly with no dilaton
derivatives has been presented already in letter \cite{NOO}.

The action (\ref{i}) diverges in general since it
contains the infinite volume integration on $M_{d+1}$.
The action is regularized by introducing the infrared cutoff
$\epsilon$ and replacing
\be
\label{vibc}
\int d^{d+1}x\rightarrow \int d^dx\int_\epsilon d\rho \ ,\ \
\int_{M_d} d^d x\Bigl(\cdots\Bigr)\rightarrow
\int d^d x\left.\Bigl(\cdots\Bigr)\right|_{\rho=\epsilon}\ .
\ee
We also expand $g_{ij}$ and $\phi$ with respect to $\rho$:
\be
\label{viib}
g_{ij}=g_{(0)ij}+\rho g_{(1)ij}+\rho^2 g_{(2)ij}+\cdots \ ,\quad
\phi=\phi_{(0)}+\rho \phi_{(1)}+\rho^2 \phi_{(2)}+\cdots \ .
\ee
Then the action is also expanded as a power series on $\rho$.
The subtraction of the terms proportional to the inverse power of
$\epsilon$ does not break the invariance under the scale
transformation $\delta g_{ \mu\nu}=2\delta\sigma g_{ \mu\nu}$ and
$\delta\epsilon=2\delta\sigma\epsilon$ . When $d$ is even, however,
the term proportional to $\ln\epsilon$ appears. This term is not
invariant under the scale transformation and the subtraction of
the $\ln\epsilon$ term breaks the invariance. The variation of the
$\ln\epsilon$ term under the scale transformation
is finite when $\epsilon\rightarrow 0$ and should be canceled
by the variation of the finite term (which does not
depend on $\epsilon$) in the action since the original action
(\ref{i}) is invariant under the scale transformation.
Therefore the $\ln\epsilon$ term $S_{\rm ln}$ gives the Weyl
anomaly $T$ of the action renormalized by the subtraction of
the terms which diverge when $\epsilon\rightarrow 0$ (d=4)
\be
\label{vib}
S_{\rm ln}=-{1 \over 2}
\int d^4x \sqrt{-g }T\ .
\ee
The conformal anomaly can be also obtained from the surface
counterterms, which is discussed in Section \ref{SS}.

First we consider the case of $d=2$, i.e. three-dimensional gauged SG.
The anomaly term $S_{\rm ln}$ proportional
to ${\rm ln}\epsilon$ in the action is
\bea
\label{IIii}
S_{\rm ln}=-{1 \over 16\pi G}{l\over 2}\int d^{2}x
\sqrt{-g_{(0)}}
\left\{ R_{(0)} + X(\phi_{(0)})(\nabla\phi_{(0)})^2 + Y(\phi_{(0)})
\Delta\phi_{(0)} \right.&&\nn
\left. + \phi _{(1)}\Phi '(\phi_{(0)})+{1 \over 2}
g^{ij}_{(0)}g_{(1)ij}\Phi(\phi_{(0)}) \right\} . &&
\eea
The terms proportional to $\rho ^{0}$ with $\mu ,\nu =i,j $
in (\ref{iii}) lead to $g_{(1)ij}$ in terms of $g_{(0)ij}$ and
$\phi_{(1)}$.
\bea
\label{vi}
g_{(1)ij}&=&\left[-R_{(0)ij}-V(\phi_{(0)})
\partial_i\phi_{(0)}\partial_j\phi_{(0)}
 -g_{(0)ij}\Phi'(\phi_{(0)})\phi_{(1)}\right.\nn
&& +{g_{(0)ij} \over l^2}\left\{2\Phi'(\phi_{(0)})\phi_{(1)}
+R_{(0)}+V(\phi_{(0)})g_{(0)}^{kl}\partial_k\phi_{(0)}
\partial_l\phi_{(0)}  \right\}\nn
&& \times \left.\left( \Phi(\phi_{(0)})
+{2 \over l^2} \right)^{-1}\right]
\times \Phi(\phi_{(0)})^{-1}
\eea
In the equation (\ref{ii}), the terms proportional to $\rho ^{-1}$
lead to $\phi_{(1)}$ as following.
\bea
\label{vii}
\phi_{(1)}&=& \left[  V'(\phi_{(0)})
g_{(0)}^{ij}\partial_i\phi_{(0)}\partial_j\phi_{(0)}
+ 2 {V(\phi_{(0)})  \over \sqrt{-g_{(0)}}} \partial_i
\left(\sqrt{-g_{(0)}}
g_{(0)}^{ij}\partial_j\phi_{(0)} \right) \right.\nn
&& \left. +{1 \over 2}\Phi'(\phi_{(0)})
\left( \Phi(\phi_{(0)})+{2 \over l^2} \right)^{-1}
\{ R_{(0)}+V(\phi_{(0)})
g^{ij}_{(0)}\partial_i\phi_{(0)}\partial_j\phi_{(0)} \} \right] \nn
&& \times \left( \Phi ''(\phi_{(0)})
 -\Phi'(\phi_{(0)})^2  \left( \Phi(\phi_{(0)})
+{2 \over l^2} \right)^{-1} \right)^{-1}
\eea
Then anomaly term takes the following form  using
(\ref{vi}), (\ref{vii})
\bea
\label{ano2}
T&=&{1 \over 8\pi G}{l\over 2}
\left\{  R_{(0)}+X(\phi_{(0)})(\nabla\phi_{(0)})^2 + Y(\phi_{(0)})
\Delta\phi_{(0)} \right. \nn
&& +{1 \over 2}\left\{ {2 \Phi '(\phi_{(0)}) \over l^2 }
\left(\Phi ''(\phi_{(0)})
\left( \Phi(\phi_{(0)})+{2 \over l^2 }\right)
 -\Phi'(\phi_{(0)})^2\right)^{-1} -\Phi(\phi_{(0)})\right\}  \nn
&& \times \left(R_{(0)} +V(\phi_{(0)})g^{ij}_{(0)}\partial_i\phi_{(0)}
\partial_j\phi_{(0)} \right)\left( \Phi(\phi_{(0)})
+{2 \over l^2 }\right)^{-1} \nn
&& +{2 \Phi '(\phi_{(0)}) \over l^2 }
\left(\Phi ''(\phi_{(0)})
\left( \Phi(\phi_{(0)})+{2 \over l^2 }\right)
 -\Phi'(\phi_{(0)})^2\right)^{-1} \nn
&& \left. \times \left(V'(\phi_{(0)})
g_{(0)}^{ij}\partial_i\phi_{(0)}\partial_j\phi_{(0)}
+ 2 {V(\phi_{(0)})  \over \sqrt{-g_{(0)}}}
\partial_i\left(\sqrt{-g_{(0)}}
g_{(0)}^{ij}\partial_j\phi_{(0)} \right) \right) \right\}\ .
\eea
For $\Phi(\phi)=0$ case, the central charge of two-dimensional conformal
field theory is defined by the coefficient of $R$. Then it
might be natural to introduce the candidate c-function
$c$ for the case when the
conformal symmetry is broken by the deformation in the
following way :
\bea
\label{d2c}
c&=&{3 \over 2G}\left[ l +{l \over 2}\left\{ {2 \Phi '(\phi_{(0)})
\over l^2 } \left(\Phi ''(\phi_{(0)})
\left( \Phi(\phi_{(0)}) \right.\right.\right.\right. \nn
&& \left.\left.\left.\left. +{2 \over l^2 }\right)
 -\Phi'(\phi_{(0)})^2\right)^{-1} -\Phi(\phi_{(0)})\right\}
\times \left( \Phi(\phi_{(0)})
+{2 \over l^2 }\right)^{-1}  \right]\ .
\eea
Comparing this with radiatively-corrected c-function of
boundary QFT (${\rm AdS_3/CFT}_2$) may help in correct
bulk description of such theory. 
Clearly, that in the regions (or for potentials) where such candidate
c-function
is singular or not monotonic it cannot be the acceptable
c-function. Presumbly, the appearence of such
regions indicates to the breaking of SG description.

Four-dimensional case is more interesting but also much more involved.
The anomaly terms which proportional to ${\rm ln }\epsilon$
are
\bea
\label{ano}
S_{\rm ln}&=&{1 \over 16\pi G}\int d^4x \sqrt{-g_{(0)}}\left[
{-1 \over 2l}g_{(0)}^{ij}g_{(0)}^{kl}\left(g_{(1)ij}g_{(1)kl}
 -g_{(1)ik}g_{(1)jl}\right) \right. \nn
&& +{l \over 2}\left(R_{(0)}^{ij}-{1 \over
2}g_{(0)}^{ij}R_{(0)}\right)g_{(1)ij} \nn
&& -{2 \over l}V(\phi_{(0)})\phi_{(1)}^2
+{l \over 2}V'(\phi_{(0)})\phi_{(1)}
g_{(0)}^{ij}\partial_i\phi_{(0)}\partial_j\phi_{(0)} \nn
&& +l V(\phi_{(0)})\phi_{(1)}
{1 \over \sqrt{-g_{(0)}}}
\partial_i\left(\sqrt{-g_{(0)}}g_{(0)}^{ij}
\partial_j\phi_{(0)} \right) \nn
&&  +{l \over 2}V(\phi_{(0)})\left( g_{(0)}^{ik}g_{(0)}^{jl}
g_{(1)kl}-{1 \over 2}g_{(0)}^{kl}
g_{(1)kl}g_{(0)}^{ij}\right)  \partial_i\phi_{(0)}
\partial_j\phi_{(0)}  \\
&& - {l \over 2}\left({1 \over 2}g_{(0)}^{ij}g_{(2)ij}
 -{1 \over 4}g_{(0)}^{ij}g_{(0)}^{kl}g_{(1)ik}g_{(1)jl}
+{1 \over 8}(g_{(0)}^{ij}g_{(1)ij})^2 \right)\Phi(\phi_{(0)})\nn
&& \left. -{l \over 2}\left( \Phi'(\phi_{(0)})\phi_{(2)}+
{1 \over 2}\Phi''(\phi_{(0)}) \phi_{(1)}^2 +
{1 \over 2}g_{(0)}^{kl}g_{(1)kl}\Phi'(\phi_{(0)}) \phi_{(1)} \right)
\right]
\ .\nonumber
\eea
The terms proportional to $\rho ^{0}$ with $\mu ,\nu =i,j $
in the equation of the motion (\ref{iii}) lead to $g_{(1)ij}$
in terms of $g_{(0)ij}$ and $\phi_{(1)}$.
\bea
\label{vibb}
g_{(1)ij}&=&\left[-R_{(0)ij}-V(\phi_{(0)})
\partial_i\phi_{(0)}\partial_j\phi_{(0)}
 -{1 \over 3}g_{(0)ij}\Phi'(\phi_{(0)})\phi_{(1)}\right.\nn
&& +{g_{(0)ij} \over l^2}\left\{{4 \over 3}\Phi'(\phi_{(0)})\phi_{(1)}
+R_{(0)}+V(\phi_{(0)})g_{(0)}^{kl}\partial_k\phi_{(0)}
\partial_l\phi_{(0)}  \right\}\nn
&& \times \left.\left( {1 \over 3}\Phi(\phi_{(0)})
+{6 \over l^2} \right)^{-1}\right]
\times \left( {1 \over 3}\Phi(\phi_{(0)})
+{2 \over l^2} \right)^{-1} \ .
\eea
In the equation (\ref{ii}), the terms proportional to $\rho^{-2}$
lead to $\phi_{(1)}$ as follows:
\bea
\label{vii4d}
\phi_{(1)}&=& \left[  V'(\phi_{(0)})
g_{(0)}^{ij}\partial_i\phi_{(0)}\partial_j\phi_{(0)}
+ 2 {V(\phi_{(0)})  \over \sqrt{-g_{(0)}}} \partial_i\left(\sqrt{-g_{(0)}}
g_{(0)}^{ij}\partial_j\phi_{(0)} \right) \right.\nn
&& \left. +{1 \over 2}\Phi'(\phi_{(0)})
\left( {1 \over 3}\Phi(\phi_{(0)})+{6 \over l^2} \right)^{-1}
\{ R_{(0)}+V(\phi_{(0)})
g^{ij}_{(0)}\partial_i\phi_{(0)}\partial_j\phi_{(0)} \} \right] \nn
&& \times \left( {8 V(\phi_{(0)}) \over l^2 } +\Phi ''(\phi_{(0)})
 -{2 \over 3}\Phi'(\phi_{(0)})^2  \left( {1 \over 3}\Phi(\phi_{(0)})
+{6 \over l^2} \right)^{-1} \right)^{-1}\ .
\eea
In the equation (\ref{iii}), the terms proportional to
$\rho^1$ with $\mu ,\nu =i,j$ lead to $g_{(2)ij}$.
\bea
\label{viii}
g_{(2)ij}&=& \left[ -{1 \over 3}\left\{ g_{(1)ij}\Phi'(\phi_{(0)})\phi_{(1)}
+g_{(0)ij}(\Phi'(\phi_{(0)})\phi_{(2)}+{1 \over 2}
\Phi''(\phi_{(0)})\phi_{(1)}^{2} ) \right\}\right.\nn
&& -{2 \over l^2}g^{kl}_{(0)}g_{(1)ki}g_{(1)lj}
+{1\over l^2}g^{km}_{(0)}g^{nl}_{(0)}g_{(1)mn}g_{(1)kl}g_{(0)ij}\nn
&& -{2 \over l^2}g_{(0)ij}\left( {1 \over 3}\Phi(\phi_{(0)})
+{8 \over l^2} \right)^{-1}\times \left\{ {2 \over l^2}
g^{mn}_{(0)}g^{kl}_{(0)}g_{(1)km}g_{(1)ln} \right.\nn
&&-{4 \over 3}\left(\Phi'(\phi_{(0)})\phi_{(2)}+{1 \over 2}
\Phi''(\phi_{(0)})\phi_{(1)}^2 \right)
 -{1 \over 3}g^{ij}_{(0)}g_{(1)ij}\Phi'(\phi_{(0)})\phi_{(1)}\nn
&& \left.+V'(\phi_{(0)})\phi_{(1)}g^{ij}_{(0)}\partial_i\phi_{(0)}
\partial_j\phi_{(0)}+{2 V(\phi_{(0)})\phi_{(1)} \over \sqrt{-g_{(0)} } }
\partial_i \left( \sqrt{-g_{(0)}}g^{ij}_{(0)}
\partial_j\phi_{(0)}\right) \right\} \nn
&& \left.+V'(\phi_{(0)})\phi_{(1)}\partial_i\phi_{(0)}
\partial_j\phi_{(0)}+2V(\phi_{(0)})\phi_{(1)}\partial_i
\partial_j\phi_{(0)} \right] \nn
&& \times \left( {1 \over 3}\Phi(\phi_{(0)}) \right)^{-1}\ .
\eea
And the terms proportional to $\rho ^{-1}$ in the equation
(\ref{ii}), lead to $\phi_{(2)}$ as follows:
\bea
\label{phi2}
\phi_{(2)}&=&\left[ V''(\phi_{(0)})\phi_{(1)}g^{ij}_{(0)}
\partial_i\phi_{(0)}\partial_j\phi_{(0)} \right.\nn
&&+V'(\phi_{(0)})\left( g^{ik}_{(0)}g^{jl}_{(0)}-{1 \over 2}
g^{ij}_{(0)}g^{kl}_{(0)}\right)g_{(1)kl}\partial_i\phi_{(0)}
\partial_j\phi_{(0)} \nn
&&+{2 V'(\phi_{(0)})\phi_{(1)} \over \sqrt{-g_{(0)} } }
\partial_i \left( \sqrt{-g_{(0)}}g^{ij}_{(0)}
\partial_j\phi_{(0)} \right) \nn
&& -{4 \over l^2}V'_{(0)}\phi_{(1)}^2-{1 \over 2}\Phi'''(\phi_{(0)})
\phi_{(1)}^2-{1 \over 2}g^{kl}_{(0)}g_{(1)kl}\Phi''(\phi_{(0)})
\phi_{(1)} \nn
&& -\left({-1 \over 4}g^{ij}_{(0)}g^{kl}_{(0)}g_{(1)ik}g_{(1)jl}
+{1 \over 8}(g^{ij}_{(0)}g_{(1)ij})^2 \right)\Phi'(\phi_{(0)}) \nn
&& -{1 \over 2}\Phi'(\phi_{(0)})\left( {1 \over 3}\Phi(\phi_{(0)})
+{8 \over l^2} \right)^{-1}\times \left\{ {2 \over l^2}
g^{mn}_{(0)}g^{kl}_{(0)}g_{(1)km}g_{(1)ln} \right.\nn
&& -{2 \over 3}\Phi''(\phi_{(0)})\phi_{(1)}^2
 -{1 \over 3}g^{ij}_{(0)}g_{(1)ij}\Phi'(\phi_{(0)})\phi_{(1)}\nn
&& \left.\left.+V'(\phi_{(0)})\phi_{(1)}g^{ij}_{(0)}\partial_i\phi_{(0)}
\partial_j\phi_{(0)}
+{2 V(\phi_{(0)})\phi_{(1)} \over \sqrt{-g_{(0)} } }
\partial_i \left( \sqrt{-g_{(0)}}g^{ij}_{(0)}
\partial_j\phi_{(0)}\right)  \right\} \right] \nn
&& \times \left( \Phi''(\phi_{(0)}) -{2 \over 3}\Phi'(\phi_{(0)})^2
\left( {1 \over 3}\Phi(\phi_{(0)})
+{8 \over l^2} \right)^{-1} \right)^{-1}
\eea
Then we can get the anomaly (\ref{ano}) in terms of
$g_{(0)ij}$ and $\phi_{(0)}$, which are boundary values of
metric and dilaton respectively by using (\ref{vibb}), (\ref{vii4d}),
(\ref{viii}), (\ref{phi2}). In the following, we choose $l=1$,
 denote $\Phi(\phi_{(0)})$ by $\Phi$ and abbreviate the
index $(0)$ for the simplicity.
Then substituting (\ref{vii4d}) into (\ref{vibb}), we obtain
\bea
\label{S1}
g_{(1)ij}&=& \tilde c_1 R_{ij} + \tilde c_2 g_{ij} R
+ \tilde c_3 g_{ij}g^{kl}\partial_{k}\phi\partial_{l}\phi \nn
&& + \tilde c_4 g_{ij}{\partial_{k} \over \sqrt{-g}}\left(
\sqrt{-g}g^{kl}\partial_{l}\phi\right)
+ \tilde c_5 \partial_{i}\phi\partial_{j}\phi\ .
\eea
The explicit form of $\tilde c_1$, $\tilde c_2$, $\cdots$
$\tilde c_5$ is given in
Appendix \ref{AA}.
Further, substituting (\ref{vii4d}) and (\ref{S1}) into
(\ref{phi2}), one gets
\bea
\label{S2}
\phi_{(2)}&=& d_1 R^2 + d_2 R_{ij}R^{ij}
+ d_3 R^{ij}\partial_{i}\phi\partial_{j}\phi \nn
&& + d_4 Rg^{ij}\partial_{i}\phi\partial_{j}\phi
+ d_5 R{1 \over \sqrt{-g}}\partial_{i}
(\sqrt{-g}g^{ij}\partial_{j}\phi) \nn
&& + d_6 (g^{ij}\partial_{i}\phi\partial_{j}\phi)^2
+ d_7 \left({1 \over \sqrt{-g}}\partial_{i}
(\sqrt{-g}g^{ij}\partial_{j}\phi)\right)^2 \nn
&& + d_8 g^{kl}\partial_{k}\phi\partial_{l}\phi
{1 \over \sqrt{-g}}\partial_{i}(\sqrt{-g}g^{ij}\partial_{j}\phi)\ .
\eea
Here, the explicit form of $d_1$, $\cdots$ $d_8$ is given in
Appendix \ref{AA}.
Substituting (\ref{vii4d}), (\ref{S1}) and (\ref{S2}) into
(\ref{viii}), one gets
\bea
\label{S3}
g^{ij}g_{(2)ij}&=& f_1 R^2 + f_2 R_{ij}R^{ij}
+ f_3 R^{ij}\partial_{i}\phi\partial_{j}\phi \nn
&& + f_4 Rg^{ij}\partial_{i}\phi\partial_{j}\phi
+ f_5 R{1 \over \sqrt{-g}}\partial_{i}
(\sqrt{-g}g^{ij}\partial_{j}\phi) \nn
&& + f_6 (g^{ij}\partial_{i}\phi\partial_{j}\phi)^2
+ f_7 \left({1 \over \sqrt{-g}}\partial_{i}
(\sqrt{-g}g^{ij}\partial_{j}\phi)\right)^2 \nn
&& + f_8 g^{kl}\partial_{k}\phi\partial_{l}\phi
{1 \over \sqrt{-g}}\partial_{i}(\sqrt{-g}g^{ij}\partial_{j}\phi) \ .
\eea
Again,
the explicit form of  very complicated functions $f_1$, $\cdots$ $f_8$
is  given in
Appendix \ref{AA}.
Finally  substituting (\ref{vii4d}), (\ref{S1}), (\ref{S2})
and (\ref{S3}) into the expression for the anomaly
(\ref{ano}), we obtain,
\bea
\label{AN1}
T&=&-{1 \over 8\pi G}\left[ h_1 R^2 + h_2 R_{ij}R^{ij}
+ h_3 R^{ij}\partial_{i}\phi\partial_{j}\phi \right. \nn
&& + h_4 Rg^{ij}\partial_{i}\phi\partial_{j}\phi
+ h_5 R{1 \over \sqrt{-g}}\partial_{i}
(\sqrt{-g}g^{ij}\partial_{j}\phi) \nn
&& + h_6 (g^{ij}\partial_{i}\phi\partial_{j}\phi)^2
+ h_7 \left({1 \over \sqrt{-g}}\partial_{i}
(\sqrt{-g}g^{ij}\partial_{j}\phi)\right)^2 \nn
&& \left. + h_8 g^{kl}\partial_{k}\phi\partial_{l}\phi
{1 \over \sqrt{-g}}\partial_{i}(\sqrt{-g}g^{ij}\partial_{j}\phi)
\right] \ .
\eea
Here
\bea
\label{h12}
h_1&=& \left[ 3\ \left\{(24-10\ \Phi)\ {\Phi'^6} \right. \right. \nn
&& + \big(62208+22464\ \Phi+2196\ {\Phi^2}+72
\ {\Phi^3}+{\Phi^4}\big)\ \Phi''\ {{(\Phi''+8\ V)}^2} \nn
&& + 2\ {\Phi'^4}\ \left\{\big(108+162\ \Phi+7\ {\Phi^2}\big)\
\Phi''+72\ \big(-8+14\ \Phi+{\Phi^2}\big)\ V\right\} \nn
&& - 2\ {\Phi'^2}\ \left\{\big(6912+2736\ \Phi+192
\ {\Phi^2}+{\Phi^3}\big)\ {\Phi''^2} \right. \nn
&& + 4\ \big(11232+6156\ \Phi+552\ {\Phi^2}
+13\ {\Phi^3}\big)\ \Phi''\ V \nn
&& \left. + 32\ \big(-2592+468\ \Phi+96\ {\Phi^2}+5
\ {\Phi^3}\big)\ {V^2}\right\} \nn
&& \left.\left. - 3\ (-24+\Phi)\ {{(6+\Phi)}^2}\ {\Phi'^3}\ (
\Phi'''+8\ V')\right\}\right] \big/  \nn
&& \left[16\ {{(6+\Phi)}^2}\ \left\{-2\ {\Phi'^2}
+(24+\Phi)\ \Phi''\right\}\ \left\{-2\ {\Phi'^2} \right.\right. \nn
&& \left.\left.+(18+\Phi)\ (\Phi''+8\ V)\right\}^2\right]\nn
h_2 &=&-\frac{3\ \left\{(12-5\ \Phi)\ {\Phi'^2}+(288+72\
\Phi+{\Phi^2})\ \Phi''\right\}}{8\ {{(6+\Phi)}^2}\
\left\{-2\ {\Phi'^2}+(24+\Phi)\ \Phi''\right\}} \ .
\eea
We also give the explicit forms of $h_3$, $\cdots$ $h_8$
in Appendix \ref{AA}. Thus, we found the complete Weyl anomaly
 from bulk side. This expression which should describe dual d4 QFT
of QCD type, with broken SUSY looks really complicated.
The interesting remark is that Weyl anomaly is not integrable in general.
In other words, it is impossible to construct the anomaly induced action.
This is not strange, as it is usual situation for conformal anomaly
when radiative corrections are taken into account.

In case of the dilaton gravity in
\cite{NOano} corresponding to $\Phi=0$ (or more generally
in case that the axion is
included \cite{GGP} as in \cite{NOOSY}), we have the following expression:
\bea
\label{Dxix}
T&=&{l^3 \over 8\pi G}\int d^4x \sqrt{-g_{(0)}}
\left[ {1 \over 8}R_{(0)ij}R_{(0)}^{ij}
 -{1 \over 24}R_{(0)}^2 \right. \nn
&& - {1 \over 2} R_{(0)}^{ij}\partial_i\varphi_{(0)}
\partial_j\varphi_{(0)} + {1 \over 6} R_{(0)}g_{(0)}^{ij}
\partial_i\varphi_{(0)}\partial_j\varphi_{(0)}  \nn
&& \left. + {1 \over 4}
\left\{{1 \over \sqrt{-g_{(0)}}} \partial_i\left(\sqrt{-g_{(0)}}
g_{(0)}^{ij}\partial_j\varphi_{(0)} \right)\right\}^2 + {1 \over 3}
\left(g_{(0)}^{ij}\partial_i\varphi_{(0)}\partial_j\varphi_{(0)}
\right)^2 \right]\ .
\eea
Here $\varphi$ can be regarded as dilaton.
In the limit of $\Phi\rightarrow 0$, we obtain
\bea
\label{Lmt}
h_1&\rightarrow& {3\cdot 62208 \Phi'' (8V)^2 \over 16\cdot 6^2
\cdot 24 \cdot 18 \Phi'' (8V)^2} = {1 \over 24} \nn
h_2&\rightarrow& - {3\cdot 288 \Phi'' \over 8\cdot 6^2\cdot 24
\Phi''}=-{1 \over 8} \nn
h_3 &\rightarrow& -{3\cdot 288 (\Phi''V - \Phi'V') \over 4\cdot 6^2 \cdot
24 \Phi''}
= - {1 \over 4}{(\Phi''V - \Phi'V') \over \Phi''} \nn
h_4 &\rightarrow& { 3\cdot 62208 \Phi'' V (8V)^2
+ 6\Phi'\cdot 384\cdot (-5184) \cdot V^2 V'
\over 8\cdot 6^2 \cdot 24 \Phi'' \cdot (18\cdot 8V)^2}
= {1 \over 12}{(\Phi''V - \Phi'V') \over \Phi''} \nn
h_5 &\rightarrow& 0 \nn
h_6 &\rightarrow& \left\{ - \Phi''\cdot 64 V \cdot \left(373248 V^3
 - 139968 {V'}^2\right) \right. \nn
&& \left. + 2\cdot 6 \Phi' V' \cdot (-2)\cdot(-432)\cdot
\left( 4608 V^3 + 864 {V'}^2 - 1728 V V''\right)\right\} \nn
&& \big/ 16\cdot 6^2 \cdot 24 \Phi'' \cdot (18\cdot 8V)^2 \nn
&=& { \left\{ -  \Phi'' V \cdot \left( V^3
 - {3 \over 8} {V'}^2\right)
+ 2 \Phi' V' \cdot \left( V^3
+ {3 \over 16} {V'}^2 - {3 \over 8} V V''\right)\right\}
\over 12 \Phi''  V^2} \nn
h_7 &\rightarrow& { V \cdot 8 \cdot 18^2 \Phi'' V \cdot 2 \cdot 12 V
\over 24 \Phi''\cdot (18\cdot 8V)^2}
= {V \over 8} \nn
h_8 &\rightarrow& {32\cdot 18^2 \Phi'' V \cdot 2\cdot 12 \cdot V'
\over 4\cdot 24 \Phi'' (18\cdot 8V)^2}
= { V'\over 8 V}\ .
\eea
Especially if we choose
\be
\label{L1}
V=-2\ ,
\ee
we obtain,
\bea
\label{L2}
&& h_1\rightarrow {1 \over 24}\ ,\quad
h_2\rightarrow -{1 \over 8}\ ,\quad
h_3 \rightarrow {1 \over 2}\ ,\quad
h_4 \rightarrow -{1 \over 6} \nn
&& h_5 \rightarrow 0 \ ,\quad h_6 \rightarrow
 -{ 1 \over 3} \ ,\quad
h_7 \rightarrow - {1 \over 4} \ ,\quad
h_8 \rightarrow 0
\eea
and we  find that the standard result (conformal
anomaly of ${\cal N}=4$ super YM theory covariantly coupled
with ${\cal N}=4$ conformal supergravity \cite{peter}) in (\ref{Dxix})
is reproduced \cite{NOano, LT}.

We should also note that the expression (\ref{AN1}) cannot be
rewritten as a sum of the Gauss-Bonnet
invariant $G$ and the square of the Weyl tensor $F$,
which are given as
\bea
\label{GF}
G&=&R^2 -4 R_{ij}R^{ij}
+ R_{ijkl}R^{ijkl} \nn
F&=&{1 \over 3}R^2 -2 R_{ij}R^{ij}
+ R_{ijkl}R^{ijkl} \ ,
\eea
This is the signal that the conformal symmetry is broken already 
in classical theory.

When $\phi$ is constant, only two terms corresponding to $h_1$
and $h_2$ survive in (\ref{AN1}) :
\bea
\label{AN1b}
T&=&-{1 \over 8\pi G}\left[ h_1 R^2 + h_2 R_{ij}R^{ij}\right] \nn
&=&-{1 \over 8\pi G}\left[ \left(h_1 + {1 \over 3}h_2\right) R^2
+ {1 \over 2}h_2\left(F-G\right)\right].
\eea
As $h_1$ depends on $V$, we
may compare the result with the conformal anomaly from, say,
scalar or spinor QED, or QCD in the phase where there are no background
scalars and (or) spinors..
The structure of the conformal anomaly in such a theory
has the following form
\be
\label{QED}
T=\hat a G + \hat b F + \hat c R^2\ .
\ee
where
\be
\label{QED1b}
\hat a=\mbox{constant} + a_1 e^2\ , \quad
\hat b=\mbox{constant}+ a_2 e^2\ , \quad
\hat c=  a_3 e^2\ .
\ee
Here $e^2$ is the electric
charge (or $g^2$ in case of QCD). Imagine that one can
identify $e$  with the exponential of the
constant dilaton (using holographic RG \cite{BK,VV}). $a_1$, $a_2$ and $a_3$
are some numbers.
Comparing (\ref{AN1b}) and (\ref{QED}), we obtain
\be
\label{QED2}
\hat a=-\hat b={h_2 \over 16\pi G}\ ,\quad
\hat c=-{1 \over 8\pi G}\left(h_1 + {1 \over 3}h_2\right) \ .
\ee
When $\Phi$ is small, one gets
\bea
\label{PP1}
h_1&=&{1 \over 24}\left[ 1 - {1 \over 8}\Phi
+ {1 \over 8}{\left(\Phi'\right)^2 \over \Phi''} \right. \nn
&& \left. + {25 \over 2592}\Phi^2 - {17 \over 216}
{\left(\Phi'\right)^2 \Phi \over \Phi''}
+ {1 \over 576}{\left(\Phi'\right)^2 \over V}
+ {1 \over 96}{\left(\Phi'\right)^4 \over \left(\Phi''\right)^2 }
+ {\cal O}\left(\Phi^3\right)\right] \nn
h_2&=&-{1 \over 8}\left[ 1 - {1 \over 8}\Phi
+ {1 \over 8}{\left(\Phi'\right)^2 \over \Phi''} \right. \nn
&& \left. + {5 \over 576}\Phi^2 - {3 \over 64}
{\left(\Phi'\right)^2 \Phi \over \Phi''}
+ {1 \over 96}{\left(\Phi'\right)^4 \over \left(\Phi''\right)^2 }
+ {\cal O}\left(\Phi^3\right)\right] \ .
\eea
If one assumes
\be
\label{PP2}
\Phi(\phi)=a\e^{b\phi}\ ,\quad (|a|\ll 1)\ ,
\ee
then
\bea
\label{PP3}
h_2&=&-{1 \over 8}\left[1 - {a^2 \over 36}\e^{2b\phi}
+ {\cal O}\left(a^3\right) \right] \nn
h_1+{1 \over 3}h_2&=&{a^2 \over 24}\left(-{5 \over 162}
+ {b^2 \over 576 V}\right)\e^{2b\phi}
+ {\cal O}\left(a^3\right) \ .
\eea
Comparing (\ref{PP3}) with (\ref{QED1b}) and (\ref{QED2})
and assuming
\be
\label{PP4}
e^2=\e^{2b\phi}\ ,
\ee
we find
\bea
\label{PP5}
a_1=-a_2&=&{1 \over 16\pi G}\cdot {1 \over 8}
\cdot {a^2 \over 36} \ , \nn
a_3&=&-{1 \over 8\pi G}\cdot {a^2 \over 24} \cdot
\left(-{5 \over 162} + {b^2 \over 576 V}\right)\ .
\eea
Here $V$ should be arbitrary but constant.
We should note $\Phi(0)\neq 0$. One can absorb the difference
into the redefinition of $l$ since we need not to assume
$\Phi(0)=0$ in deriving the form of $h_1$ and $h_2$ in
(\ref{h12}). Hence, this simple example suggests the way of comparison
between SG side and QFT descriptions of non-conformal boundary theory.


In order that the region near the boundary at $\rho=0$ is
asymptotically AdS, we need to require $\Phi\rightarrow 0$
and $\Phi'\rightarrow 0$ when $\rho \rightarrow 0$.
One can also confirm that $h_1\rightarrow {1 \over 24}$ and
$h_2\rightarrow -{1 \over 8}$ in the limit of $\Phi\rightarrow 0$
and $\Phi'\rightarrow 0$
even if $\Phi''\neq 0$ and $\Phi'''\neq 0$.
In the AdS/CFT correspondence, $h_1$ and $h_2$ are related with
the central charge $c$ of the conformal field theory
(or its analog for non-conformal theory). Since
we have two functions $h_1$ and $h_2$, there are two ways to define
the candidate c-function when the conformal field theory is deformed:
\be
\label{CC}
c_1={24\pi h_1 \over G}\ ,\quad
c_2=-{8\pi h_2 \over G}\ .
\ee
If we put $V(\phi)=4\lambda^2 + \Phi(\phi)$, then
$l=\left(12\over V(0)\right)^{1 \over 2}$. One should note that
it is chosen $l=1$  in (\ref{CC}). We can
restore $l$ by changing $h\rightarrow l^3 h$ and $k\rightarrow
l^3 k$ and $\Phi'\rightarrow l\Phi'$, $\Phi''\rightarrow
l^2\Phi''$ and $\Phi''' \rightarrow l^3\Phi'''$ in (\ref{AN1}).
Then in the limit of $\Phi\rightarrow 0$, one gets
\be
\label{CCl}
c_1\ ,\quad c_2\ \rightarrow {\pi \over G}
\left(12\over V(0)\right)^{3 \over 2}\ ,
\ee
which agrees with the proposal of the previous
work \cite{GPPZ2} in the limit.
The c-function $c_1$ or $c_2$ in (\ref{CC}) is, of course,
more general definition.
It is interesting to study the behaviour of candidate c-function for
explicit values of dilatonic potential at
different limits. It also could be interesting to see
what is the analogue of our dilaton-dependent
c-function in non-commutative YM theory
(without dilaton, see \cite{wu}).

\section{Properties of c-function}

The definitions of the c-functions in (\ref{d2c}) and (\ref{CC}),
are, however, not always good ones since our results are too wide.
That is, we have obtained the conformal anomaly for arbitrary
dilatonic background which may not be the solution of original
$d=5$ gauged supergravity. As only solutions of d5 gauged supergravity
describe RG flows of dual QFT it is not strange that above candidate
c-functions are not acceptable. They quickly become non-monotonic
and even singular in explicit examples. They presumbly measure the
deviations from SG description and should not be taken seriously.
As pointed in \cite{MTR}, it might be necessary to impose the
condition $\Phi'=0$ on the conformal boundary. Such condition follows
from the equations of motion of d5 gauged SG.
Anyway as $\Phi'= 0$ on the boundary in the solution which has
the asymptotic AdS region, we can add any function
which proportional to the power of $\Phi'= 0$ to the previous
expressions of the c-functions in (\ref{d2c}) and (\ref{CC}).
As a trial, if we put $\Phi'=0$, we obtain
\bea
\label{d2cb}
c&=&{3 \over 2G}\left[ {l \over 2} + {1 \over l}
{1 \over \Phi(\phi_{(0)})  +{2 \over l^2 }}\right]
\eea
instead of (\ref{d2c}) and
\bea
\label{CCb}
c_1&=&{2\pi \over 3G}{62208+22464\Phi
+2196 \Phi^2 +72 \Phi^3+ \Phi^4 \over
(6+\Phi)^2(24+\Phi)(18+\Phi)} \nn
c_2&=&{3\pi \over G}{288+72 \Phi+ \Phi^2 \over
(6+\Phi)^2(24+\Phi)}
\eea
instead of (\ref{CC}).
We should note that there disappear the higher
derivative terms like $\Phi''$ or $\Phi'''$. That will be our
final proposal for acceptable c-function in terms of dilatonic potential.
The given c-functions in (\ref{CCb}) also have the property
(\ref{CCl}) and reproduce the known result for the central charge
on the boundary.
Since ${d\Phi \over dz}\rightarrow 0$ in the asymptotically AdS
region even if the region is UV or IR, the given c-functions in
(\ref{d2cb}) and (\ref{CCb}) have fixed points in the
asymptotic AdS region ${d c \over dU}={dc \over d\Phi}
{d\Phi \over d\phi}{d\phi \over dU}\rightarrow 0$, where
$U=\rho^{-{1 \over 2}}$ is the radius coordinate in AdS
or the energy scale of the boundary field theory.

We can now check the monotonity in the c-functions.
For this purpose, we consider some examples.
In \cite{FGPW} and \cite{GPPZ}, the
following dilaton potentials appeared:
\bea
\label{FGPWpot}
4\lambda^2 + \Phi_{\rm FGPW}(\phi)
&=&4\left(\exp\left[ \left({4\phi \over \sqrt{6}}\right)
\right] + 2 \exp\left[ -\left({2\phi \over \sqrt{6}}\right)
\right]\right)\\
\label{GPPZpot}
4\lambda^2 + \Phi_{\rm GPPZ}(\phi)
&=&{3 \over 2}\left(3+\left(\cosh\left[ \left({2\phi \over
\sqrt{3}}\right)\right]\right)^2 + 4\cosh\left[ \left(
{2\phi \over \sqrt{3}}\right)\right]\right) \ .
\eea
In both cases $V$ is a constant and $V=-2$.
In the classical solutions for the both cases,
$\phi$ is the monotonically decreasing
function of the energy scale $U= \rho^{-{1 \over 2}}$ and
$\phi=0$ at the UV limit corresponding
to the boundary.
Then in order to know the energy scale dependences
of $c_1$ and $c_2$, we only need to investigate the $\phi$
dependences of $c_1$ and $c_2$ in (\ref{CCb}). As the potentials
and also $\Phi$ have a minimum $\Phi=0$
at $\phi=0$, which corresponds to the UV boundary in the solutions
in \cite{FGPW} and \cite{GPPZ}, and $\Phi$ is monotonicaly
increasing function of the absolute value $|\phi|$, we only
need to check the monotonities of $c_1$ and $c_2$ with respect
to $\Phi$ when $\Phi\geq 0$. From (\ref{CCb}), we find
\bea
\label{monot}
&& {d \left(\ln c_1\right) \over d\Phi} \nn
&& = - {18\left(622080 + 383616\Phi + 64296\Phi^2 + 4548\Phi^3
+ 130\Phi^4 + \Phi^5\right) \over
(6 + \Phi)(18 + \Phi) (24 + \Phi) (62208+22464\Phi
+2196 \Phi^2 +72 \Phi^3+ \Phi^4 )} \nn
&& <0 \nn
&& {d \left(\ln c_2\right) \over d\Phi}=
 - {5184 + 2304\Phi + 138\Phi^2 + \Phi^3
 \over (6 + \Phi) (24 + \Phi) (288+72 \Phi+ \Phi^2)}
<0 \ .
\eea
Therefore the c-functions $c_1$ and $c_2$ are monotonically
decreasing functions of $\Phi$ or increasings function of the
energy scale $U$ as the c-function in \cite{DF,GPPZ}.
We should also note that the
c-functions $c_1$ and $c_2$ are positive definite for
non-negative $\Phi$.  For $c$ in (\ref{d2cb}) for $d=2$
case, it is very straightforward to check the monotonity and the
positivity.

In \cite{GPPZ2}, another c-function has been proposed
in terms of the metric as follows:
\be
\label{gppzC}
c_{\rm GPPZ}=\left({dA \over d z}\right)^{-3}\ ,
\ee
where the metric is given by
\be
\label{gppzC2}
ds^2=dz^2 + \e^{2A}dx_\mu dx^\mu\ .
\ee
The c-function (\ref{gppzC}) is positive and
has a fixed point in the
asymptotically AdS region again and the c-function is also
 monotonically increasing function of the energy scale.
The c-functions (\ref{d2cb}) and (\ref{CCb}) proposed
in this paper are given in terms of the dilaton potential,
not in terms of metric, but it might be interesting that
the c-functions in (\ref{d2cb}) and (\ref{CCb}) have the
similar properties (positivity, monotonity and fixed point
in the asymptotically AdS region).
These properties could be understood from the equations of motion.
When the metric has the form (\ref{gppzC2}), the equations
of motion are:
\bea
\label{Ei}
&& \phi''+ dA'\phi' = {\partial \Phi \over \partial \phi}\ , \\
\label{Eii}
&& d A''+ d (A')^2 + {1 \over 2}(\phi')^2
= - {4\lambda^2 + \Phi \over d-1} \ , \\
\label{Eiii}
&& A'' + d (A')^2 = - {4\lambda^2 + \Phi \over d-1} \ .
\eea
Here $'\equiv {d \over dz}$. From (\ref{Ei}) and (\ref{Eii}),
we obtain
\be
\label{E1}
 0=2(d-1)A'' + {\phi'}^2
\ee
Then if $A''=0$, $\phi'=0$, which tells that
if ${d c_{\rm GPPZ} \over dz}=0$, then ${dc_1 \over dz}={dc_2
\over dz}=0$. Then $c_{\rm GPPZ}$ has a fixed point, $c_1$ and $c_2$
have a fixed point. From (\ref{Ei}) and (\ref{Eii}),
we also obtain
\be
\label{E2}
 0=d(d-1){A'}^2 +4\lambda^2 + \Phi  - {1 \over 2}{\phi'}^2\ .
\ee
Then at the fixed point where $\phi'=0$, we obtain
\be
\label{E2b}
 0=d(d-1){A'}^2 + 4\lambda^2 + \Phi \ .
\ee
Therefore if $c_{\rm GPPZ}$ and $A'$ is the monotonic function
of $z$, $V$ and also $c_1$ and $c_2$ are also monotonic function
at least at the fixed point. We have to note that above considerations 
do not give the proof of equivalency of our proposal c-functions with
other proposals. However, it is remarkable (at least, for a number of 
potentials)
that they enjoy the similar properties: positivity, monotonity and 
existance of fixed points.

We can also consider other examples of c-function for
different choices of dilatonic potential.
In \cite{CLP}, several examples of the potentials in gauged
supergravity are given. They appeared as a result of sphere
reduction in M-theory or string theory, down to three or five
dimensions. Their properties are
described in detail in refs.\cite{CLP}. The potentials have
the following form:
\be
\label{pot}
4\lambda^2 + \Phi(\phi) = {d(d-1) \over {1 \over a_1^2}
 - {1 \over a_1 a_2}}\left( {1 \over a_1^2}\e^{a_1 \phi}
 - {1 \over a_1 a_2}\e^{a_2\phi} \right) \ .
\ee
Here $a_1$ and $a_2$ are constant parameters depending on the
model. We also normalize the potential so that
$4\lambda^2 + \Phi(\phi) \rightarrow d(d-1)$ when
$\phi\rightarrow 0$. For simplicity, we choose $G=l=1$ in
this section.

For ${\cal N}=1$ model in $D=d+1=3$ dimensions
\be
\label{d21}
a_1=2\sqrt{2}\ ,\quad a_2=\sqrt{2}\ ,
\ee
for $D=3$, ${\cal N}=2$, one gets
\be
\label{d22}
a_1=\sqrt{6}\ ,\quad a_2=2\sqrt{2 \over 3}\ ,
\ee
and for $D=3$, ${\cal N}=3$ model, we have
\be
\label{d23}
a_1={4 \over \sqrt{3}}\ ,\quad a_2=\sqrt{3}\ .
\ee
On the other hand, for $D=d+1=5$, ${\cal N}=1$ model, $a_1$
and $a_2$ are
\be
\label{d41}
a_1=2\sqrt{5 \over 3}\ ,\quad a_2={4 \over \sqrt{15}}\ .
\ee
The proposed c-functions have not acceptable behaviour for above potentials.
  (There seems to be  no problem for 2d case.)
The problem seems to be that the
solutions in above models have not asymptotic AdS region
in UV but in IR. On the same time the conformal anomaly in (\ref{AN1}) is
evaluated as UV effect. If we assume that $\Phi$ in the expression
of c-functions $c_1$ and $c_2$ vanishes at IR AdS region,
$\Phi$ becomes negative. When $\Phi$ is negative, the properties
of the c-functions $c_1$ and $c_2$ become bad, they are not
monotonic nor positive, and furthermore they have a singularity
in the region given by the solutions in \cite{CLP}.
Thus, for such type of potential other proposal for c-function 
which isnot related with conformal nomaly should be made.

Hence, we discussed the typical behaviour of candidate c-functions.
However, it is not clear which role should play dilaton
in above expressions as holographic RG coupling constant in dual QFT.
It could be induced mass, quantum fields or coupling constants (most
probably, gauge 
coupling),
but the explicit
rule with what it should be identified is absent.
The big number of usual RG parameters in dual QFT
suggests also that there should be considered gauged SG
with few scalars.

\section{Surface Counterterms and Finite Action \label{SS}}

As well-known, we need to add the surface terms
to the bulk action in order to have the well-defined variational principle.
 Under the variation of the metric
$\hat G^{\mu\nu}$ and the scalar field $\phi$, the variation of
the action (\ref{i}) is given by
\bea
\label{Ii}
\lefteqn{\delta S=\delta S_{M_{d+1}} + \delta S_{M_d}} \\
\lefteqn{\delta S_{M_{d+1}}={1 \over 16\pi G}
\int_{M_{d+1}} d^{d+1} x \sqrt{-\hat G}\left[
\delta \hat G^{\zeta\xi}\left\{-{1 \over 2}G_{\zeta\xi}\left\{
\hat R \right.\right.\right.} \nn
&& \left.\left. + \left(X(\phi) - Y'(\phi)\right)(\hat\nabla\phi)^2
+ \Phi (\phi)+4 \lambda ^2 \right\}
+ \hat R_{\zeta\xi}+
\left(X(\phi) - Y'(\phi)\right)\partial_\zeta\phi
\partial_\xi\phi \right\} \nn
&& + \delta\phi\left\{\left(X'(\phi) - Y''(\phi)\right)
(\hat\nabla\phi)^2 + \Phi' (\phi) \right. \nn
&& \left.\left. - {1 \over \sqrt{-\hat G}}\partial_\mu\left(
\sqrt{-\hat G}\hat G^{\mu\nu}\left(X(\phi) - Y'(\phi)\right)
\partial_\nu\phi\right)\right\}\right]. \nn
\lefteqn{\delta S_{M_d}= {1 \over 16\pi G}
\int_{M_d} d^d x \sqrt{-\hat g}n_\mu\left[
\partial^\mu
\left(\hat G_{\xi\nu}\delta \hat G^{\xi\nu}\right)
 - D_\nu \left(\delta \hat G^{\mu\nu}\right)
 + Y(\phi)\partial^\mu\left(\delta\phi\right)
\right]\ .}\nonumber
\eea
Here $\hat g_{\mu\nu}$ is the metric induced from
$\hat G_{\mu\nu}$ and $n_\mu$ is the unit vector normal to
$M_d$.
The surface term $\delta S_{M_d}$ of the variation contains
$n^\mu\partial_\mu\left(\delta\hat G^{\xi\nu}\right)$ and
$n^\mu\partial_\mu\left(\delta\phi\right)$, which makes the
variational principle ill-defined.
In order that the variational principle is well-defined on the
boundary, the variation of the action should be written as
\be
\label{bdry1}
\delta S_{M_d}=\lim_{\rho\rightarrow 0}
\int_{M_d} d^d x \sqrt{-\hat g}\left[\delta\hat G^{\xi\nu}
\left\{\cdots\right\} + \delta\phi\left\{\cdots\right\}\right]
\ee
after using the partial integration. If we put
$\left\{\cdots\right\}=0$ for $\left\{\cdots\right\}$ in
(\ref{bdry1}), one could obtain the boundary condition
corresponding to Neumann boundary condition. We can, of course, 
select Dirichlet boundary condition by choosing 
$\delta\hat G^{\xi\nu}=\delta\phi=0$, which is natural for 
AdS/CFT correspondence. The Neumann type condition becomes, 
however, necessary later when we consider the black hole 
mass etc. by using surface terms. 
If the variation of the action on the boundary contains
$n^\mu\partial_\mu\left(\delta\hat G^{\xi\nu}\right)$ or
$n^\mu\partial_\mu\left(\delta\phi\right)$, however, we
cannot partially integrate it on the boundary in order to rewrite
the variation in the form of (\ref{bdry1}) since
$n_\mu$ expresses the direction perpendicular
to the boundary. Therefore the ``minimum'' of the action is
ambiguous. Such a problem was well studied in \cite{3} for the
Einstein gravity and the boundary term was added to the action.
It cancels the term containing
$n^\mu\partial_\mu\left(\delta\hat G^{\xi\nu}\right)$. We need
to cancel also the term containing
$n^\mu\partial_\mu\left(\delta\phi\right)$. Then one finds the
boundary term  \cite{NOano}
\be
\label{bdry2}
S_b^{(1)} = -{1 \over 8\pi G}
\int_{M_d} d^d x \sqrt{-\hat g} \left[D_\mu n^\mu
+ Y(\phi)n_\mu\partial^\mu\phi\right]\ .
\ee
We also need to add surface counterterm $S_b^{(2)}$ which cancels
the divergence coming from the infinite volume of the bulk space,
say AdS. In order to investigate the divergence,
we choose the metric in the form (\ref{ib}).
In the parametrization (\ref{ib}), $n^\mu$ and the curvature
$R$ are given by
\bea
\label{Iiii}
n^\mu&=&\left({2\rho \over l},0,\cdots,0\right) \nn
R&=&\tilde R + {3\rho^2 \over l^2}\hat g^{ij} \hat g^{kl}
\hat g_{ik}' \hat g_{jl}' - {4\rho^2 \over l^2}
\hat g^{ij} \hat g_{ij}''
 - {\rho^2 \over l^2}\hat g^{ij} \hat g^{kl}
\hat g_{ij}' \hat g_{kl}' \ .
\eea
Here $\tilde R$ is the scalar curvature defined by $g_{ij}$ 
in (\ref{ib}). 
Expanding $g_{ij}$ and $\phi$ with respect to $\rho$ as
in (\ref{viib}),
we find the following expression for $S+S_b^{(1)}$:
\bea
\label{II}
S+S_b^{(1)}&=&{1 \over 16\pi G}\lim_{\rho\rightarrow 0}
\int d^d x l \rho^{-{d \over 2}}
\sqrt{-g_{(0)}}\left[ {2-2d \over l^2}
 - {1 \over d}\Phi(\phi_0)\right. \nn
&& + \rho\left\{ -{1 \over d-2} R_{(0)}
 - {1 \over l^2}g_{(0)}^{ij} g_{(1)ij} \right. \nn
&& \left. - {1 \over d-2} \right( X(\phi_{(0)}) \left(\nabla_{(0)}
\phi_{(0)}\right)^2 + Y(\phi_{(0)})\Delta \phi_{(0)} \nn
&& \left.\left. + \Phi'(\phi_{(0)})\phi_{(1)} \Biggr) \right\} +
{\cal O}\left(\rho^2\right) \right] \ .
\eea
Then  for $d=2$
\be
\label{III}
S_b^{(2)}={1 \over 16\pi G}
\int d^d x \sqrt{-\hat g}\left[ {2 \over l}
+ {l \over 2}\Phi(\phi)\right]
\ee
and for $d=3,4$,
\bea
\label{IIIb}
&& S_b^{(2)}={1 \over 16\pi G}
\int d^d x \left[\sqrt{-\hat g}\left\{ {2d-2 \over l}
+ {l \over d-2}R - {2l \over d(d-2)}\Phi(\phi)\right.\right.  \nn
&& \left. + {l \over d-2}\left(X(\phi)
\left(\hat\nabla\phi\right)^2 + Y(\phi)\hat\Delta\phi\right)
\right\}\left. - {l^2 \over d(d-2)}n^\mu\partial_\mu
\left(\sqrt{-\hat g}\Phi(\phi)\right)\right]\ .
\eea
Note that the last term in above expression does not look typical 
from the AdS/CFT point of view. The reason is that it does not 
depend from only the boundary values of the fields. Its presence 
may indicate to breaking of AdS/CFT conjecture in the situations 
when SUGRA scalars significally deviate from  constants or are 
not asymptotic constants 
\footnote{We thank the referee for adressing this issue.}. 

Here 
$\hat\Delta$ and $\hat\nabla$ are defined by using $d$-dimensional 
metric and 
we used
\bea
\label{IV}
&& \sqrt{-\hat g}\Phi(\phi)=\rho^{-{d \over 2}}
\sqrt{-g_{(0)}}\Biggl\{\Phi(\phi_{(0)}) \nn
&& \quad \left.
+ \rho\left({1 \over 2}g_{(0)}^{ij}g_{(1)ij}\Phi(\phi_{(0)})
+ \Phi'(\phi_{(0)})\phi_{(1)} \right)
+ {\cal O}\left(\rho^2\right)\right\} \nn
&& n^\mu\partial_\mu\left(\sqrt{-\hat g}\Phi(\phi)\right)
={2 \over l}\rho^{-{d \over 2}}
\sqrt{-g_{(0)}}\left\{-{d \over 2}\Phi(\phi_{(0)} \right.\nn
&& \left. \ \qquad + \rho\left(1 - {d \over 2}\right)
\left({1 \over 2}g_{(0)}^{ij}g_{(1)ij}\Phi(\phi_{(0)})
+ \Phi'(\phi_{(0)})\phi_{(1)}\right) + {\cal O}\left(\rho^2\right)
\right\} \ .
\eea
Note that $S_b^{(2)}$ in (\ref{III}) or (\ref{IIIb}) is
only given in terms of the boundary quantities except
the last term in (\ref{IIIb}). The last term is
necessary to cancel the divergence of the bulk action
and it is, of course, the total derivative
in the bulk theory:
\be
\label{V}
\int d^d x
n^\mu\partial_\mu\left(\sqrt{-\hat g}\Phi(\phi)\right)
=\int d^{d+1}x\sqrt{-\hat G}\Box\Phi(\phi)\ .
\ee
 Thus we got the boundary counterterm action for gauged SG. Using these
local surface counterterms as part of complete action one can
show explicitly that bosonic sector of gauged SG in dimensions under
discussion gives finite action in asymptotically AdS space.
The corresponding example will be given in next section.

Recently the surface counterterms for the action with the
dilaton (scalar) potential are discussed in \cite{CO}. Their
counterterms seem to correspond to the terms cancelling the
leading divergence when $\rho\rightarrow 0$ in (\ref{II}). However,
they seem to have only considered the case where the dilaton becomes
asymptotically constant $\phi\rightarrow \phi_0$. If we choose
$\phi_0=0$, the total dilaton potential including the cosmological
term $V_{\rm dilaton}(\phi)\equiv 4\lambda^2 + \Phi(\phi)$
approaches to $V_{\rm dilaton}(\phi)\rightarrow 4\lambda^2
= d(d-1)/{l^{2}}$. Then if we only consider the leading $\rho$
behavior and the asymptotically constant dilaton,
the counterterm action in (\ref{III}) and/or (\ref{IIIb}) has
the following form
\bea
\label{IIIc}
S_b^{(2)}&=&{1 \over 16\pi G}
\int d^d x \sqrt{-\hat g}\left( {2d-2 \over l} \right)\ ,
\eea
which coincides with the result in \cite{CO} when the
spacetime is aymptotically AdS.

Let us turn now to the discussion of deep connection between surface
counterterms and holographic conformal anomaly.
It is enough to mention only $d=4$.
In order to control the logarithmically divergent terms in the
bulk action $S$, we choose $d-4=\epsilon<0$. Then
\be
\label{D4SS}
S+S_b={1 \over \epsilon}S_{\ln} +\ \mbox{finite terms}\ .
\ee
Here $S_{\ln}$ is given in (\ref{ano}).
We also find
\be
\label{Eeq}
g_{(0)}^{ij}{\delta \over \delta g_{(0)}^{ij}} S_{\ln}
=-{\epsilon \over 2}{\cal L}_{\ln} + {\cal O}
\left(\epsilon^2\right)\ .
\ee
Here ${\cal L}_{\ln}$ is the Lagrangian density
corresponding to $S_{\ln}$ : $S_{\ln}=\int d^{d+1}
{\cal L}_{\ln}$.
Then combining (\ref{D4SS}) and (\ref{Eeq}),
we obtain the trace anomaly :
\be
\label{Sx}
T=\lim_{\epsilon \rightarrow 0-}
{2\hat g_{(0)}^{ij} \over \sqrt{- \hat g_{(0)}}}{\delta (S+S_b)
\over \delta \hat g_{(0)}^{ij}}
=-{1 \over 2}{\cal L}_{\ln}\ ,
\ee
which is identical with the result found in (\ref{vib}). 
We should note that the last term in (\ref{IIIb}) does not 
lead to any ambiguity in the calculation of conformal anomaly since
$g_{(0)}$ does not depend on 
$\rho$. 
If we use the equations of motion (\ref{vibb}), (\ref{vii4d}),
(\ref{viii}) and (\ref{phi2}), we finally obtain the
expression  (\ref{AN1}) or (\ref{AN1A}).
Hence, we found the finite gravitational action
(for asymptotically AdS spaces) in 5 dimensions
by adding the local surface counterterm. This action
correctly reproduces  holographic trace anomaly for
dual (gauge) theory. In principle, one can also generalize
all results for higher dimensions, say, d6, etc. With
the growth of dimension, the technical problems become
more and more complicated as the number of structures
in boundary term is increasing.

\section{Dilatonic AdS Black Hole and its Mass}

Let us consider the black hole or ``throat'' type solution for the
equations of the motion (\ref{ii}) and (\ref{iii}) when
$d=4$. The surface term (\ref{IIIb}) may be used for calculation
of the finite black hole mass
 and/or other thermodynamical quantities.

For simplicity, we choose
\be
\label{BHi}
X(\phi)=\alpha\ (\mbox{constant})\ ,\quad
Y(\phi)=0\
\ee
and we assume the spacetime metric in the following form:
\be
\label{BHii}
ds^2=-\e^{2\rho}dt^2 + \e^{2\sigma}dr^2 + r^2\sum_{i=1}^{d-1}
\left(dx^i\right)^2
\ee
and $\rho$, $\sigma$ and $\phi$  depend only on $r$.
The equations (\ref{ii}) and (\ref{iii}) can be rewritten in
the following form:
\bea
\label{BHiii}
0&=&\e^{\rho+\sigma}\Phi'(\phi) - 2\alpha \left(\e^{\rho-\sigma}
\phi'\right)' \\
\label{BHiv}
0&=& -{1 \over 3}\e^{2\rho}\left(\Phi(\phi)+{12 \over l^2}\right)
+ \left(\rho'' + \left(\rho'\right)^2 - \rho'\sigma' +
{3\rho' \over r}\right)\e^{2\rho-2\sigma} \\
\label{BHv}
0&=& {1 \over 3}\e^{2\sigma}\left(\Phi(\phi)+{12 \over l^2}\right)
 - \rho'' - \left(\rho'\right)^2 + \rho'\sigma' +
{3\sigma' \over r} + \alpha\left(\phi'\right)^2 \\
\label{BHvi}
0&=& {1 \over 3}\e^{2\sigma}\left(\Phi(\phi)+{12 \over l^2}\right)
r^2 + k + \left\{r\left(\sigma' - \rho'\right) -2 \right\}
\e^{-2\sigma}\ .
\eea
Here $'\equiv {d \over dr}$. If one defines new variables $U$ and
$V$ by
\be
\label{BHvii}
U=\e^{\rho+\sigma}\ ,\quad V=r^2\e^{\rho-\sigma}\ ,
\ee
we obtain the following equations from (\ref{BHiii}-\ref{BHvi}):
\bea
\label{BHviii}
0&=&r^3 U\Phi'(\phi)-2\alpha\left(rV\phi'\right)' \\
\label{BHix}
0&=& {1 \over 3}\e^{2\sigma}\left(\Phi(\phi)+{12 \over l^2}\right)
r^3U + kr - V' \\
\label{BHx}
0&=& {3U' \over rU} + \alpha\left(\phi\right)'\ .
\eea
We should note that only three equations in (\ref{BHiii}-\ref{BHvi})
are independent. There is practical problem in the construction of AdS BH
with non-trivial dilaton, especially for arbitrary dilatonic potential.
That is why we use below the approximate technique which was developed in
ref.\cite{NOR} for constant dilatonic potential.

When $\Phi(0)=\Phi'(0)=\phi=0$, a solution corresponding to the
throat limit of D3-brane is given by
\be
\label{BHxi}
U=1\ ,\quad V=V_0\equiv {r^4 \over l^2} - \mu\ .
\ee
In the following, we use large $r$ expansion and consider the
perturbation around (\ref{BHxi}). It is assumed
\be
\label{BHxii}
\Phi(\phi)=\tilde\mu \phi^2 + {\cal O}\left(\phi^3\right)\ .
\ee
Then one can neglect the higher order terms in (\ref{BHxii}).
We obtain from (\ref{BHviii})
\be
\label{BHxiii}
0\sim \tilde\mu r^3 \phi + \alpha \left({r^5 \over l^2}\phi'
\right)'\ .
\ee
The solution of eq.(\ref{BHxiii}) is given by
\be
\label{BHxiv}
\phi=cr^{-\beta}\ ,\ (\mbox{$c$ is a constant})\ ,
\beta=2\pm \sqrt{4-{\tilde\mu l^2 \over \alpha}}\ .
\ee
Consider $r$ is large or $c$ is small, and write $U$ and $V$ in the
following form:
\be
\label{BHxv}
U=1+c^2 u \ ,\quad V=V_0+ c^2 v\ .
\ee
Then from (\ref{BHix}) and(\ref{BHx}), one gets
\be
\label{BHxvi}
u=u_0+{\alpha\beta \over 6}r^{-2\beta}\ ,
\quad v= v_0 - {\tilde\mu (\beta -6) \over 6(\beta -4)(\beta -2)}
r^{-2\beta +4}\ .
\ee
Here $u_0$ and $v_0$ are constants of the integration.
Here we choose
\be
\label{BHxvii}
v_0=u_0=0\ .
\ee
The horizon which is defined by
\be
\label{BHxviii}
V=0
\ee
lies at
\be
\label{BHxix}
r=r_h\equiv l^{1 \over 2}\mu^{1 \over 4} + c^2
{\tilde\mu (\beta -6)l^{{5 \over 2}-\beta}
\mu^{{1 \over 4} - {\beta \over 2}} \over
24(\beta -4)(\beta -2)}\ .
\ee
And the Hawking temperature is
\bea
\label{BHxx}
T&=&{1 \over 4\pi}\left[{1 \over r^2}
{d V \over dr}\right]_{r=r_h} \nn
&=& {1 \over 4\pi}\left\{4l^{-{3 \over 2}}\mu^{1 \over 4}
+ c^2{\tilde\mu (\beta -6)(2\beta - 3) \over 6(\beta -4)(\beta -2)}
l^{{1 \over 2} - \beta} \mu^{{1 \over 4} - {\beta \over 2}}
\right\}\ .
\eea

We now evaluate the free energy of the black hole within the
standard prescription \cite{GKT,GKP}.
The free energy $F$ can be obtained by substituting
the classical solution into the action $S$:
\be
\label{BHxxb}
F=TS\ .
\ee
Here $T$ is the Hawking temperature.
 Using the equations of motion in (\ref{ii})
($X=\alpha$, $Y=0$, $4\lambda^2={12 \over l^2}$),
we obtain
\be
\label{BHxxi}
0={5 \over 3}\left(\Phi(\phi) + {12 \over l^2}\right) + \hat R
+\alpha \left(\nabla\phi\right)^2 \ .
\ee
Substituting (\ref{BHxxi}) into the action (\ref{i}) after
Wick-rotating it to the Euclid signature
\bea
\label{BHxxii}
S&=&{1 \over 16\pi G}\cdot {2 \over 3}\int_{M_5}d^5 \sqrt{G}
\left(\Phi(\phi) + {12 \over l^2}\right) \nn
&=&{1 \over 16\pi G}\cdot {2 \over 3}
{V_{(3)} \over T} \int_{r_h}^\infty dr r^3 U
\left(\Phi(\phi) + {12 \over l^2}\right) \ .
\eea
Here $V_{(3)}$ is the volume of the 3d space
($\int d^5 x \cdots = \beta V_{(3)} \int dr r^3 \cdots$)
and $\beta$ is the period of time, which can be regarded as
the inverse of the temperature $T$ (${1 \over T}$).
The expression (\ref{BHxxii}) contains the divergence. We
regularize the divergence by replacing
\be
\label{BHxxiii}
\int^\infty dr \rightarrow \int^{r_{\rm max}} dr
\ee
and subtract the contribution from a zero temperature solution,
where we choose $\mu=c=0$, and the solution corresponds to the
vacuum or
pure AdS:
\be
\label{BHxxiv}
S_0={1 \over 16\pi G}\cdot {2 \over 3}\cdot {12 \over l^2}
{V_{(3)} \over T} \sqrt{G_{tt}\left(r=r_{\rm max}, \mu=c=0\right)
\over G_{tt}\left(r=r_{\rm max}\right) }
\int_{r_h}^\infty dr r^3 \ .
\ee
The factor $\sqrt{G_{tt}\left(r=r_{\rm max}, \mu=c=0\right)
\over G_{tt}\left(r=r_{\rm max}\right) }$ is chosen so that the
proper length of the circles  which correspond to the
period ${1 \over T}$ in the Euclid time at $r_{\rm max}$
coincides with each other in the two solutions.
Then we find the following expression for the free energy,
\bea
\label{BHxxv}
F&=&\lim_{r_{\rm max}\rightarrow \infty}
T\left(S - S_0\right) \nn
&=&{V_{(3)} \over 2\pi G l^2 T^2}\left[ - {l^2 \mu \over 8}
+ c^2 \mu^{1-{\beta \over 2}}\tilde \mu \left\{
{(\beta -1) \over 12 \beta (\beta - 4)(\beta -2)}
\right\}+ \cdots \right] \ .
\eea
Here we assume $\beta > 2$ or the expression  $S - S_0$ still
contains the divergences and we cannot get finite results.
However, the inequality $\beta > 2$ is not always satisfied
in the gauged supergravity models. In that case the expression
in (\ref{BHxxv}) would not be valid.
 One can express the free energy $F$ in
(\ref{BHxxv}) in terms of the temperature $T$ instead of $\mu$:
\be
\label{BHxxvi}
F={V_{(3)} \over 16\pi G}\left[ -\pi T^4 l^6 + c^2l^{8-4\beta}
T^{4-2\beta}\tilde\mu \left({2\beta^3 - 15 \beta^2 + 22\beta - 4
\over 6\beta(\beta - 4)(\beta -2)}\right) + \cdots \right]\ .
\ee
Then the entropy ${\cal S}$ and the energy (mass) $E$
is given by
\bea
\label{BHxxvii}
{\cal S}&=&-{dF \over dT}
={V_{(3)} \over 16\pi G}\Bigl[ 4\pi T^3 l^6  \nn
&& + c^2l^{8-4\beta}
T^{3-2\beta}\tilde\mu \left({2\beta^3 - 15 \beta^2 + 22\beta - 4
\over 3\beta(\beta - 4)}\right) + \cdots \Bigr] \nn
E&=&F+TS ={V_{(3)} \over 16\pi G}\Bigl[ 3\pi T^4 l^6  \nn
&& + c^2l^{8-4\beta}
\left(\pi T^4\right)^{1-{\beta \over 2}}\tilde\mu
\left({(2\beta - 3)(2\beta^3 - 15 \beta^2 + 22\beta - 4)
\over 6\beta(\beta - 4)(\beta -2)}\right) + \cdots \Bigr]\ .
\eea

We now evaluate the mass  using the surface term of the action
in (\ref{IIIb}), i.e. within local surface counterterm method.
 The surface energy momentum tensor $T_{ij}$
is now defined by ($d=4$)\footnote{ 
$S$ does not contribute due to the 
equation of motion in the bulk. 
The variation of $S+S_b^{(1)}$ gives a 
contribution proportional to the extrinsic curvature 
$\theta_{ij}$ at the boundary:
\[
\delta \left(S + S_b^{(2)}\right)=
{\sqrt{-\hat g} \over 16\pi G }\left(\theta_{ij} 
 - \theta \hat g_{ij}
\right)\delta\hat g^{ij}
\] 
The contribution is finite even in the limit of 
$r\rightarrow \infty$. Then the finite part does not 
depend on the parameters characterizing the black hole. 
Therefore after subtracting the contribution from the 
reference metric, which could be that of AdS, the 
contribution from the variation of $S+S_b^{(1)}$ vanishes.}
\bea
\label{BHxxviii}
\delta S_b^{(2)}&=&
\sqrt{-\hat g}\delta\hat g^{ij}T_{ij} \nn
&=& {1 \over 16\pi G}\left[\sqrt{-\hat g}\delta\hat g^{ij}
\left\{-{1 \over 2}\hat g_{ij}
\left({6 \over l} + {l \over 2}\hat R
+ {l \over 4}\Phi(\phi)\right)\right\} \right. \nn
&& \left. + {l^2 \over 4} n^\mu \partial_\mu\left\{
\sqrt{-\hat g}\delta\hat g^{ij} \hat g_{ij}\Phi(\phi)\right\}
\right]\ .
\eea
Note that the energy-momentum tensor is still not
well-defined due to the term containing $ n^\mu\partial_\mu$.
If we assume $\delta\hat g^{ij} \sim
{\cal O}\left(\rho^{a_1}\right)$ for large $\rho$
when we choose the coordinate system (\ref{ib}), then
\be
\label{BHxxix}
n^\mu\partial_\mu\left(\delta\hat g^{ij} \cdot \right)
\sim {2 \over l}\delta\hat g^{ij}\left(a_1
+ \partial_\rho\right) (\cdot)\ .
\ee
Or  if $\delta\hat g^{ij} \sim
{\cal O}\left(r^{a_2}\right)$ for large $r$
when we choose the coordinate system (\ref{BHii}), then
\be
\label{BHxxx}
n^\mu\partial_\mu\left(\delta\hat g^{ij} \cdot \right)
\sim \delta\hat g^{ij}\e^\sigma \left({a_2 \over r}
+\partial_r\right)
(\cdot)\ .
\ee
As we consider the black hole-like object in this section,
one chooses the coordinate system (\ref{BHii}) and assumes
Eq.(\ref{BHxxx}). Then mass $E$ of the black hole like object
is given by
\be
\label{Mii}
E=\int d^{d-1}x \sqrt{\tilde\sigma} N \delta T_{tt}
\left(u^t\right)^2\ .
\ee
Here 
we assume the metric of the 
reference spacetime (e.g. AdS) has the form of 
$ds^2 = f(r)dr^2- N^2(r)dt^2 
+ \sum_{i,j=1}^{d-1}\tilde\sigma_{ij}dx^i dx^j$ and 
$\delta T_{tt}$ is the difference of the $(t,t)$ component
of the energy-momentum tensor
in the spacetime with black hole like object from that in the
reference spacetime, which we choose to be AdS,
and $u^t$ is the $t$ component of the unit
time-like vector normal to the hypersurface given
by $t=$constant.
By using the solution in (\ref{BHxv}) and (\ref{BHxvi}),
the $(t,t)$ component of the energy-momentum tensor
in (\ref{BHxxviii}) has the following form:
\bea
\label{BHxxxi}
T_{tt}&=&{3 \over 16\pi G}{r^2 \over l^3}\left[ 1
 - {l^3 \mu \over r^4}
+ l^2\tilde\mu c^2 \left(
{1 \over 12} - {1 \over 6\beta(\beta - 6 )} \right.\right. \nn
&& \left. \left. - {\beta - 6 \over 6(\beta - 4)(\beta -2)}
 - {(3 - \beta)(1 + a_2) \over 12}
\right)r^{-2\beta} + \cdots \right]\ .
\eea
If we assume the mass is finite, $\beta$ should satisfy the
inequality $\beta > 2$, as in the case of the free energy
in (\ref{BHxxv}) since $\sqrt{\sigma} N
\left(u^t\right)^2 = lr^2$ for the reference AdS space.
Then the $\beta$-dependent term in (\ref{BHxxxi}) does not
contribute to the mass and one gets
\be
\label{BHxxxii}
E={3\mu V_{(3)} \over 16\pi G}\ .
\ee
Using (\ref{BHxx})
\be
\label{BHxxxiii}
E={3l^6 V_{(3)} \pi T^4\over 16\pi G}\left\{
1 - c^2\tilde \mu l^{2-4\beta}\left(\pi T^4\right)^{-{\beta \over 2}}
{(\beta -6) (2\beta - 3) \over
(\beta - 4)(\beta -2)}
\right\}\ ,
\ee
which does not agree with the result in (\ref{BHxxvii}).
This might express the ambiguity in the choice of the
regularization to make the finite action.
A possible origin of it might be following. We assumed $\phi$ can be
expanded in the (integer) power series of $\rho$ in
(\ref{viib}) when deriving the surface terms in
(\ref{IIIb}). However,  this assumption seems to conflict with
the classical solution in (\ref{BHxiv}), where the
fractional power seems to appear since $r^2
\sim {1 \over \rho}$.  In any case, in QFT there is no problem in
regularization dependence of the results.
In many cases (see example in ref.\cite{SNO}) the explicit choice
of free parameters of regularization
leads to coincidence of the answers which look different in different
regularizations. As usually happens in QFT the  renormalization
is more universal as the same answers for beta-functions may be
obtained while using different regularizations. That suggests
that holographic renormalization group should be
developed and the predictions of above calculations should be
tested in it.

As in the case of the c-function, we might be drop the terms
containing $\Phi'$ in the expression of $S_b^{(2)}$
in (\ref{IIIb}). Then we obtain
\bea
\label{IIIbb}
&& S_b^{(2)}={1 \over 16\pi G}
\int d^d x \left[\sqrt{-\hat g}\left\{ {2d-2 \over l}
+ {l \over d-2}R
+ {2l \over d(d-2)}\Phi(\phi)\right.\right.  \nn
&& \left. + {l \over d-2}\left(X(\phi)
\left(\hat\nabla\phi\right)^2 + Y(\phi)\hat\Delta\phi\right)
\right\}\left. - {l^2\Phi(\phi) \over d(d-2)}n^\mu\partial_\mu
\left(\sqrt{-\hat g}\right)\right]\ .
\eea
If we use the expression (\ref{IIIbb}), however, the result of
the mass $E$ in (\ref{BHxxxiii}) does not change.

\section{Discussion}

In summary, we constructed surface counterterm for gauged
supergravity
with single scalar and arbitrary scalar potential in three and five
dimensions. As a result, the finite gravitational action
and consistent stress tensor in asymptotically AdS space
is found. Using this action, the regularized expressions
for free energy, entropy and mass are derived for d5
dilatonic AdS black hole. From another side, finite action may
be used to get the holographic conformal anomaly
of boundary QFT with broken conformal invariance.
Such conformal anomaly is calculated from d5 and d3 gauged SG
with arbitrary dilatonic potential with the use of AdS/CFT
correspondence. Due to dilaton dependence it takes extremely
complicated form. Within holographic RG where identification of
dilaton with some coupling constant is made, we suggested the
candidate c-function for d2 and d4 boundary QFT from holographic 
conformal anomaly.
It is shown that such proposal gives monotonic and positive c-function 
for few examples of dilatonic potential.

We expect that our results may be very useful in explicit
identification of supergravity description (special RG flow) with the
particular
boundary gauge theory (or its phase) which is very non-trivial
task in AdS/CFT correspondence. We show that on the example
of constant dilaton and special form of dilatonic potential
where qualitative agreement of holographic conformal anomaly and
QFT conformal anomaly (with the account of radiative
corrections) from QED-like theory with single coupling constant
may be achieved.

Our work may be extended in various directions. First of all,
we can consider large number of scalars, say 42 as in $N=8$ d5 SG,
and construct the corresponding Weyl anomaly from the bulk side.
However, this is technically very complicated problem as even
in case of single scalar the complete answer for d4 anomaly
takes few pages. The calculation of surface counterterm
in d5 gauged SG with many scalars is slightly easier task.
However, again the application of surface counterterm
for the derivation of regularized thermodynamical quantities
in multi-scalar
AdS black holes (when they will be constructed) is complicated.
Second, the generalization of surface counterterm for higher
dimensions (say, $d=7,9$) is possible.
Third, in general the extension of AdS/CFT set-up to non-conformal
boundary theories is challenging problem. In this respect, better
investigation
of candidate c-functions from bulk and from boundary
is required as well as their comparison in all detail.
The related question is bulk calculation of Casimir effect
in the presence of dilaton and comparison of it with QFT result,
including radiative corrections.

\section*{Acknoweledgements}

The work by SDO has been supported in part by CONACyT (CP, ref.990356) 
and in part by RFBR grant N99-02-16617.

\appendix

\section{Coefficients of conformal anomaly\label{AA}}

In this appendix, we give the explicit values of the
coefficients appeared in the calculation of
$d=4$ conformal anomaly.

 Substituting (\ref{vii4d}) into (\ref{vibb}), we obtain
\bea
\label{S1A}
g_{(1)ij}&=& \tilde c_1 R_{ij} + \tilde c_2 g_{ij} R
+ \tilde c_3 g_{ij}g^{kl}\partial_{k}\phi\partial_{l}\phi \nn
&& + \tilde c_4 g_{ij}{\partial_{k} \over \sqrt{-g}}\left(
\sqrt{-g}g^{kl}\partial_{l}\phi\right)
+ \tilde c_5 \partial_{i}\phi\partial_{j}\phi
\\
\tilde c_1&=& -\frac{3}{6+\Phi} \nn
\tilde c_2&=& -\frac{3\ \left\{{\Phi'^2}-6
\ (\Phi''+8\ V)\right\}}{2\ (6+\Phi)
\ \left\{-2\ {\Phi'^2}+(18+\Phi)\ (\Phi'' +8\ V)\right\}} \nn
\tilde c_3&=& \frac{-3\ {\Phi'^2}\ V+18\ V\ (\Phi''+8\ V)-2\
(6+\Phi)\ \Phi'\ V'}{2\ (6+\Phi)\ (-2\ {\Phi'^2}+(18+\Phi)\
(\Phi''+8\ V))} \nn
\tilde c_4&=& -\frac{2\ \Phi'\ V}{-2\ {\Phi'^2}+(18+\Phi)\
(\Phi''+8\ V)} \nn
\tilde c_5&=& -\frac{V}{2+\frac{\Phi}{3}}\ .
\eea
Further, substituting (\ref{vii4d}) and (\ref{S1A}) into
(\ref{phi2}), we obtain
\bea
\label{S2A}
\phi_{(2)}&=& d_1 R^2 + d_2 R_{ij}R^{ij}
+ d_3 R^{ij}\partial_{i}\phi\partial_{j}\phi \nn
&& + d_4 Rg^{ij}\partial_{i}\phi\partial_{j}\phi
+ d_5 R{1 \over \sqrt{-g}}\partial_{i}
(\sqrt{-g}g^{ij}\partial_{j}\phi) \nn
&& + d_6 (g^{ij}\partial_{i}\phi\partial_{j}\phi)^2
+ d_7 \left({1 \over \sqrt{-g}}\partial_{i}
(\sqrt{-g}g^{ij}\partial_{j}\phi)\right)^2 \nn
&& + d_8 g^{kl}\partial_{k}\phi\partial_{l}\phi
{1 \over \sqrt{-g}}\partial_{i}(\sqrt{-g}g^{ij}\partial_{j}\phi) \\
d_1&=&-\left[9\ \Phi'\ \left\{2\
(12+\Phi)\ {\Phi'^4}-\big(-864+36\ \Phi+24
\ {\Phi^2}+{\Phi^3}\big)\ {\Phi''^2}  \right.\right. \nn
&& + 192\ {{(12+\Phi)}^2}\ \Phi''\ V
+64\ \big(2592+612\ \Phi+48\ {\Phi^2}+{\Phi^3}\big)\ {V^2}  \nn
&& - 2\ {\Phi'^2}\ \left(\big(216+30\ \Phi
+{\Phi^2}\big)\ \Phi''+144\ (10+\Phi)\ V\right) \nn
&& \left.\left. + {{(6+\Phi)}^2}\ (24+\Phi)\ \Phi'\ (\Phi'''+8\ V')
\right\}\right]\big/  \nn
&& \left[8\ {{(6+\Phi)}^2}\ \left\{
 -2\ {\Phi'^2}+(24+\Phi)\ \Phi''\right\}
\ \right. \nn
&& \left. \times
{{\left\{-2\ {\Phi'^2}+(18+\Phi)\ (\Phi''+8\ V)\right\}}^2}
\right] \nn
d_2&=&\frac{9\ (12+\Phi)\ \Phi'}
{4\ {{(6+\Phi)}^2}\ \left\{-2\ {\Phi'^2}+(24+\Phi)
\ \Phi''\right\}} \nn
d_3&=&\frac{3\ (3\ (12+\Phi)\ \Phi'\ V-2\ (144+30\
\Phi+{\Phi^2})\ V')}{2\ {{(6+\Phi)}^2}\ (-2\ {\Phi'^2}+(24+\Phi)
\ \Phi'')} \nn
d_4 &=&\big(3\ \big(-6\ (12+\Phi)\ {\Phi'^5}\ V+6\ \big(108+24
\ \Phi+{\Phi^2}\big)\ {\Phi'^4}\ V' \nn
&& + 4\ \big(2592+684\ \Phi+48\ {\Phi^2}+{\Phi^3}\big)\ (\Phi''+8\ V)
 \ ((9+\Phi)\ \Phi'' \nn
&& +4\ (12+\Phi)\ V)\ V' -(6+\Phi)\ {\Phi'^2}\ \big(3\ \big(144+30
\ \Phi+{\Phi^2}\big)\ \Phi'''\ V \nn
&& + \big(1980\ \Phi''+216\ \Phi\ \Phi''+5\ {\Phi^2}\ \Phi''+27360
\ V+4176\ \Phi\ V \nn
&& +128\ {\Phi^2}\ V\big)\ V'\big)
+ 2\ {\Phi'^3}\ \big(3\ \big(216+30\ \Phi+{\Phi^2}\big)
\ \Phi''\ V \nn
&& - 2\ \big(-2160\ {V^2}-216\ \Phi\ {V^2}+864\ V''+324
\ \Phi\ V'' \nn
&& +36\ {\Phi^2}\ V''+{\Phi^3}\ V''\big)\big)
+ \Phi'\ \big(3\ \big(-864+36\ \Phi+24\ {\Phi^2}+{\Phi^3}\big)
\ {\Phi''^2}\ V \nn
&& +2\ \Phi''\ \big(-41472\ {V^2}-6912
\ \Phi\ {V^2} - \ 288\ {\Phi^2}\ {V^2}+15552\ V'' \nn
&& +6696\ \Phi\ V''+972\ {\Phi^2}
\ V''+54\ {\Phi^3}\ V''+{\Phi^4}\ V''\big) \nn
&& -2\ \big(248832\ {V^3}+58752\ \Phi\ {V^3}+4608\ {\Phi^2}\ {V^3} \nn
&& + 96\ {\Phi^3}\ {V^3}+15552\ \Phi'''\ V'+6696\ \Phi\ \Phi'''\ V'
+972\ {\Phi^2}\ \Phi'''\ V' \nn
&& + 54\ {\Phi^3}\ \Phi'''\ V'+{\Phi^4}\ \Phi'''\ V'
+124416\ {V'^2}+53568\ \Phi\ {V'^2} \nn
&& + 7776\ {\Phi^2}\ {V'^2}+432\ {\Phi^3}\ {V'^2}+8\ {\Phi^4}
  \ {V'^2}-124416\ V\ V'' \nn
&& - 53568\ \Phi\ V\ V''-7776\ {\Phi^2}\ V\ V''-432\ {\Phi^3}
  \ V\ V'' \nn
&& -8\ {\Phi^4}\ V\ V''\big)\big)\big)\big)\big/  \nn
&& \big(4\ {{(6+\Phi)}^2}\ \big(-2\ {\Phi'^2}+(24+\Phi)
\ \Phi''\big)\ \big(-2\ {\Phi'^2} \nn
&& +(18+\Phi)\ (\Phi''+8\ V)\big)^2 \big) \nn
d_5 &=& -\big(3\ \big(2\ {\Phi'^4}\ V+2\ \big(432+42\ \Phi+{\Phi^2}\big)
\ \Phi''\ V\ (\Phi''+8\ V) \nn
&& + {\Phi'^2}\ V\ ((6+\Phi)\ \Phi''-8\ (162+7\ \Phi)\ V)
- 4\ (24+\Phi)\ {\Phi'^3}\ V' \nn
&&-2\ \big(432+42\ \Phi+{\Phi^2}\big)
  \ \Phi'\ (\Phi'''\ V-\Phi''\ V')\big)\big)\big/  \nn
&& \big(2\ \big(2\ {\Phi'^2}-(24+\Phi)\ \Phi''\big)\ {{\big(-2
  \ {\Phi'^2}+(18+\Phi)\ (\Phi''+8\ V)\big)}^2}\big) \nn
d_6 &=& -\big(-54\ (12+\Phi)\ {\Phi'^5}\ {V^2}
+ 12\ \big(828+168\ \Phi+5\ {\Phi^2}\big)\ {\Phi'^4}\ V
  \ V' \nn
&& +4\ \big(2592+684\ \Phi+48\ {\Phi^2}+{\Phi^3}\big)\   \nn
&& V'\ \big(54\ {\Phi''^2}\ V+4608\ {V^3}+192\ \Phi
  \ {V^3}+108\ \Phi'''\ V'+24\ \Phi\ \Phi'''\ V' \nn
&& + {\Phi^2}\ \Phi'''\ V'+864\ {V'^2}+192\ \Phi\ {V'^2}+8
  \ {\Phi^2}\ {V'^2}-1728\ V\ V'' \nn
&& -384\ \Phi\ V\ V''
 - 16\ {\Phi^2}\ V\ V''+2\ \Phi''\ \big(504\ {V^2}+12
  \ \Phi\ {V^2}-108\ V''\nn
&& -24\ \Phi\ V''-{\Phi^2}\ V''\big)\big)
+ (6+\Phi)\ {\Phi'^2}\ \big(9\ \big(144+30\ \Phi+{\Phi^2}
\big)\ \Phi'''\ {V^2} \nn
&& - 2\ V'\ \big(14796\ \Phi''\ V+1368\ \Phi\ \Phi''\ V+33\ {\Phi^2}
  \ \Phi''\ V \nn
&& +88992\ {V^2}+4680\ \Phi\ {V^2}
+ 36\ {\Phi^2}\ {V^2}-20736\ V''\nn
&& -5472\ \Phi\ V''-384
  \ {\Phi^2}\ V''-8\ {\Phi^3}\ V''\big)\big) \nn
&& + 2\ {\Phi'^3}\ \big(27\ \big(216+30\ \Phi+{\Phi^2}\big)
  \ \Phi''\ {V^2}+4\ \big(12312\ {V^3} \nn
&& + 1836\ \Phi\ {V^3}+72\ {\Phi^2}\ {V^3}+2376
  \ {V'^2}+864\ \Phi\ {V'^2}+90\ {\Phi^2}\ {V'^2} \nn
&& + 2\ {\Phi^3}\ {V'^2}+2592\ V\ V''+972\ \Phi\ V
  \ V''+108\ {\Phi^2}\ V\ V'' \nn
&& +3\ {\Phi^3}\ V\ V''\big)\big)
 - \Phi'\ \big(27\ \big(2304+516\ \Phi+40
  \ {\Phi^2}+{\Phi^3}\big)\ {\Phi''^2}\ {V^2} \nn
&& + 4\ \Phi''\ \big(217728\ {V^3}+44064\ \Phi
  \ {V^3}+3024\ {\Phi^2}\ {V^3}+72\ {\Phi^3}\ {V^3} \nn
&& +81648\ {V'^2}
+ 34992\ \Phi\ {V'^2}+5040\ {\Phi^2}\ {V'^2}+276
  \ {\Phi^3}\ {V'^2} \nn
&& +5\ {\Phi^4}\ {V'^2}+46656\ V\ V''
+ 20088\ \Phi\ V\ V'' \nn
&& +2916\ {\Phi^2}\ V\ V''+162\ {\Phi^3}\ V
  \ V''+3\ {\Phi^4}\ V\ V''\big) \nn
&& + 4\ V\ \big(746496\ {V^3}+129600\ \Phi
  \ {V^3}+6912\ {\Phi^2}\ {V^3} \nn
&& + 144\ {\Phi^3}\ {V^3}-46656\ \Phi'''\ V'-20088\ \Phi\ \Phi'''
  \ V'-2916\ {\Phi^2}\ \Phi'''\ V' \nn
&& - 162\ {\Phi^3}\ \Phi'''\ V'-3\ {\Phi^4}\ \Phi'''\ V'-404352
  \ {V'^2}-177984\ \Phi\ {V'^2} \nn
&& - 26784\ {\Phi^2}\ {V'^2}-1584\ {\Phi^3}\ {V'^2}-32\ {\Phi^4}
  \ {V'^2}+373248\ V\ V'' \nn
&& + 160704\ \Phi\ V\ V''+23328\ {\Phi^2}\ V\ V''+1296\ {\Phi^3}\ V
  \ V'' \nn
&& +24\ {\Phi^4}\ V\ V''\big)\big)\big)\big/
\big(8\ {{(6+\Phi)}^2}\ \big(-2\ {\Phi'^2}+(24+\Phi)
\ \Phi''\big)\ \big(-2\ {\Phi'^2} \nn
&& +(18+\Phi) \ (\Phi''+8\ V)\big)^2\big) \nn
d_7 &=& \big(2\ V\ \big(36\ {\Phi'^3}\ V-3
\ (18+\Phi)\ \Phi'\ V\   \nn
&& ((26+\Phi)\ \Phi''-8\ (18+\Phi)\ V)+4\ \big(432+42
  \ \Phi+{\Phi^2}\big)\ {\Phi'^2}\ V' \nn
&& + {{(18+\Phi)}^2}\ (24+\Phi)\ (\Phi'''\ V-2
  \ (\Phi''+4\ V)\ V')\big)\big)\big/  \nn
&& \big(\big(2\ {\Phi'^2}-(24+\Phi)\ \Phi''\big)
  \ {{\big(-2\ {\Phi'^2}+(18+\Phi)\ (\Phi''+8\ V)\big)}^2}\big) \nn
d_8 &=& -\big(6\ {\Phi'^4}\ {V^2}-4\ (156+5\ \Phi)\ {\Phi'^3}
\ V\ V'-2\ (18+\Phi)\ \Phi'\ V\   \nn
&& (3\ (24+\Phi)\ \Phi'''\ V+(-276\ \Phi''-11\ \Phi\ \Phi''+480
  \ V+32\ \Phi\ V)\ V') \nn
&& + 2\ \big(432+42\ \Phi+{\Phi^2}\big)\ \big(3\ {\Phi''^2}
  \ {V^2} \nn
&& +2\ (18+\Phi)\ V\ (-\Phi'''\ V'+8\ V\ V'') \nn
&& + 2\ \Phi''\ \big(12\ {V^3}+18\ {V'^2}+\Phi\ {V'^2}+18
  \ V\ V''+\Phi\ V\ V''\big)\big) \nn
&& + {\Phi'^2}\ \big(3\ (6+\Phi)\ \Phi''\ {V^2}-8
  \ \big(486\ {V^3}+21\ \Phi\ {V^3}+432\ {V'^2} \nn
&& + 42\ \Phi\ {V'^2}+{\Phi^2}\ {V'^2}+432\ V\ V''+42\ \Phi
  \ V\ V''+{\Phi^2}\ V\ V''\big)\big)\big)\big/  \nn
&& \big(2\ \big(2\ {\Phi'^2}-(24+\Phi)\ \Phi''\big)
  \ {{\big(-2\ {\Phi'^2}+(18+\Phi)\ (\Phi''+8\ V)\big)}^2}\big)\ .
\nonumber
\eea
Substituting (\ref{vii4d}), (\ref{S1A}) and (\ref{S2A}) into
(\ref{viii}), one gets
\bea
\label{S3A}
g^{ij}g_{(2)ij}&=& f_1 R^2 + f_2 R_{ij}R^{ij}
+ f_3 R^{ij}\partial_{i}\phi\partial_{j}\phi \nn
&& + f_4 Rg^{ij}\partial_{i}\phi\partial_{j}\phi
+ f_5 R{1 \over \sqrt{-g}}\partial_{i}
(\sqrt{-g}g^{ij}\partial_{j}\phi) \nn
&& + f_6 (g^{ij}\partial_{i}\phi\partial_{j}\phi)^2
+ f_7 \left({1 \over \sqrt{-g}}\partial_{i}
(\sqrt{-g}g^{ij}\partial_{j}\phi)\right)^2 \nn
&& + f_8 g^{kl}\partial_{k}\phi\partial_{l}\phi
{1 \over \sqrt{-g}}\partial_{i}(\sqrt{-g}g^{ij}\partial_{j}\phi) \\
f_1&=& \left[9\ \left\{2\ {\Phi'^6}-72\ (12+\Phi)\ \Phi''
\ {{(\Phi''+8\ V)}^2} \right.\right.\nn
&& - 2\ {\Phi'^4}\ \left((24+\Phi)\ \Phi''
+8\ (18+\Phi)\ V\right) \nn
&& +{\Phi'^2}\ \left(\big(324+12\
\Phi-{\Phi^2}\big)\ {\Phi''^2} \right. \nn
&& \left. + 8\ \big(540+48\ \Phi+{\Phi^2}\big)\ \Phi''\ V
+64\ \big(180+24\ \Phi+{\Phi^2}\big)\ {V^2}\right)  \nn
&& \left.\left. + {{(6+\Phi)}^2}\
{\Phi'^3}\ (\Phi'''+8\ V')\right\}\right]\big/  \nn
&& \left[2\ {{(6+\Phi)}^2}\ \left\{-2\ {\Phi'^2}
+(24+\Phi)\ \Phi''\right\}\ \right. \nn
&& \left. \times {{\left\{-2\ {\Phi'^2}
+(18+\Phi)\ (\Phi''+8\ V)\right\}}^2}\right] \nn
f_2&=& -\frac{9\ ({\Phi'^2}-6\ \Phi'')}{{{(6+\Phi)}^2}
\ \left\{-2\ {\Phi'^2}+(24+\Phi)\ \Phi''\right\}} \nn
f_3 &=& \frac{6\ (-3\ {\Phi'^2}\ V+18\ \Phi''\ V+2\ (6+\Phi)
\ \Phi'\ V')}{{{(6+\Phi)}^2}\ (-2\ {\Phi'^2}+(24+\Phi)
\ \Phi'')} \nn
f_4 &=& -\big(3\ \big(-12\ {\Phi'^6}\ V+432\ (12+\Phi)\ \Phi''
 \ V\ {{(\Phi''+8\ V)}^2} \nn
&& + 8\ (6+\Phi)\ {\Phi'^5}\ V'+(6+\Phi)\ \Phi'
\ \big(\big(1044+168\ \Phi+7\ {\Phi^2}\big)\ {\Phi''^2} \nn
&& + 8\ \big(1476+192\ \Phi+7\ {\Phi^2}\big)\ \Phi''
  \ V \nn
&& +256\ \big(216+30\ \Phi+{\Phi^2}\big)\ {V^2}\big)\ V' \nn
&& - 2\ (6+\Phi)\ {\Phi'^3}\ (3\ (6+\Phi)\ \Phi'''\ V \nn
&& +(66\ \Phi''+3
\ \Phi\ \Phi''+912\ V+88\ \Phi\ V)\ V') \nn
&& + 4\ {\Phi'^4}\ \big(3\ (24+\Phi)\ \Phi''\ V-2
\ \big(-216\ {V^2}-12\ \Phi\ {V^2}+36\ V'' \nn
&& +12\ \Phi\ V''+{\Phi^2}\ V''\big)\big)
+ 2\ {\Phi'^2}\ \big(3\ \big(-324-12
\ \Phi+{\Phi^2}\big)\ {\Phi''^2}\ V \nn
&& + 2\ (18+\Phi)\ \Phi''\ \big(-360\ {V^2}-12\ \Phi\ {V^2}+36
  \ V''+12\ \Phi\ V''+{\Phi^2}\ V''\big) \nn
&& - 2\ \big(17280\ {V^3}+2304\ \Phi\ {V^3} \nn
&& + 96\ {\Phi^2}\ {V^3}+648\ \Phi'''\ V'+252\ \Phi\ \Phi'''
  \ V'+30\ {\Phi^2}\ \Phi'''\ V' \nn
&& + {\Phi^3}\ \Phi'''\ V'+5184\ {V'^2}+2016\ \Phi\ {V'^2}+240
  \ {\Phi^2}\ {V'^2}+8\ {\Phi^3}\ {V'^2} \nn
&& - 5184\ V\ V''-2016\ \Phi\ V\ V''-240\ {\Phi^2}\ V
  \ V''-8\ {\Phi^3}\ V\ V''\big)\big)\big)\big)\big/  \nn
&& \big(2\ {{(6+\Phi)}^2}\ \big(-2\ {\Phi'^2}+(24+\Phi)
\ \Phi''\big)\ {{\big(-2\ {\Phi'^2}+(18+\Phi)
\ (\Phi''+8\ V)\big)}^2}\big) \nn
f_5 &=& -\big(3\ \Phi'\ \big(\Phi''\ V\ (-3\ (10+\Phi)
\ \Phi''+8\ (42+\Phi)\ V) \nn
&& + {\Phi'^2}\ \big(-6\ \Phi''\ V+32\ {V^2}\big)
+8\ {\Phi'^3}\ V' \nn
&& +4 \ (18+\Phi)\ \Phi'
\ (\Phi'''\ V-\Phi''\ V')\big)\big)\big/  \nn
&& \big(\big(2\ {\Phi'^2}-(24+\Phi)\ \Phi''\big)\ {{\big(-2
\ {\Phi'^2}+(18+\Phi)\ (\Phi''+8\ V)\big)}^2}\big) \nn
f_6 &=& \big(-54\ {\Phi'^6}\ {V^2}+72\ (6+\Phi)
\ {\Phi'^5}\ V\ V'\nn
&& +2\ \Phi''\big(54\ \big(252
+30\ \Phi+{\Phi^2}\big)\ {\Phi''^2}\ {V^2} \nn
&& +24\ V\ \big(36288\ {V^3}+4320\ \Phi\ {V^3}
+144\ {\Phi^2}\ {V^3} \nn
&& + 11664\ {V'^2}+5184\ \Phi\ {V'^2}+792\ {\Phi^2}
  \ {V'^2}+48\ {\Phi^3}\ {V'^2}+{\Phi^4}\ {V'^2}\big) \nn
&& + \Phi''\ \big(217728\ {V^3}+25920\ \Phi\ {V^3}+864
\ {\Phi^2}\ {V^3} \nn
&& + 11664\ {V'^2}+5184\ \Phi\ {V'^2}+792
\ {\Phi^2}\ {V'^2}+48\ {\Phi^3}\ {V'^2}+{\Phi^4}\ {V'^2}\big)\big) \nn
&& + (6+\Phi)\ {\Phi'^3}\ \big(9\ (6+\Phi)\ \Phi'''\ {V^2}-2\ V'
  \ \big(666\ \Phi''\ V+39\ \Phi\ \Phi''\ V \nn
&& + 4392\ {V^2}+156\ \Phi\ {V^2}-864\ V''-192\ \Phi
  \ V''-8\ {\Phi^2}\ V''\big)\big) \nn
&& + (6+\Phi)\ \Phi'\ V'\ \big(3\ \big(1548+120
  \ \Phi+{\Phi^2}\big)\ {\Phi''^2}\ V \nn
&& +8\ \Phi''\ \big(11124\ {V^2}+1152\ \Phi\ {V^2} \nn
&& + 27\ {\Phi^2}\ {V^2}-1944\ V''-540\ \Phi\ V'' \nn
&& -42\ {\Phi^2}\ V''-{\Phi^3}\ V''\big)+4\ (18+\Phi)\   \nn
&& \big(4608\ {V^3}+192\ \Phi\ {V^3}+108 \ \Phi'''\ V' \nn
&& +24\ \Phi\ \Phi'''\ V'+{\Phi^2}  \ \Phi'''\ V'+864\ {V'^2} \nn
&& + 192\ \Phi\ {V'^2}+8\ {\Phi^2}\ {V'^2}-1728\ V
  \ V''-384\ \Phi\ V\ V''-16\ {\Phi^2}\ V\ V''\big)\big) \nn
&& + 6\ {\Phi'^4}\ \big(9\ (24+\Phi)\ \Phi''\ {V^2}+4\ \big(324
  \ {V^3}+18\ \Phi\ {V^3} \nn
&& + 36\ {V'^2}+12\ \Phi\ {V'^2}+{\Phi^2}\ {V'^2}+36\ V
  \ V''+12\ \Phi\ V\ V''+{\Phi^2}\ V\ V''\big)\big) \nn
&& - {\Phi'^2}\ \big(27\ \big(396+36\ \Phi+{\Phi^2}\big)
  \ {\Phi''^2}\ {V^2}+4\ \Phi''\ \big(29160\ {V^3} \nn
&& + 2592\ \Phi\ {V^3}+54\ {\Phi^2}\ {V^3}+4104
  \ {V'^2}+1620\ \Phi\ {V'^2}+198\ {\Phi^2}\ {V'^2} \nn
&& + 7\ {\Phi^3}\ {V'^2}+1944\ V\ V''+756\ \Phi\ V\ V''+90
  \ {\Phi^2}\ V\ V''+3\ {\Phi^3}\ V\ V''\big) \nn
&& + 4\ V\ \big(67392\ {V^3}+6912\ \Phi\ {V^3}+144
  \ {\Phi^2}\ {V^3} \nn
&& -1944\ \Phi'''\ V'-756\ \Phi\ \Phi'''\ V' \nn
&& - 90\ {\Phi^2}\ \Phi'''\ V'-3\ {\Phi^3}\ \Phi'''\ V'-5184
  \ {V'^2}-2016\ \Phi\ {V'^2}-240\ {\Phi^2}\ {V'^2} \nn
&& - 8\ {\Phi^3}\ {V'^2}+15552\ V\ V''+6048\ \Phi\ V
\ V'' \nn
&& +720\ {\Phi^2}\ V\ V''+24\ {\Phi^3}\ V\ V''\big)\big)\big)\big/  \nn
&& \big(2\ {{(6+\Phi)}^2}\ \big(-2\ {\Phi'^2}+(24+\Phi)
\ \Phi''\big)\ \big(-2\ {\Phi'^2} \nn
&& +(18+\Phi)\ (\Phi''+8\ V) \big)^2\big) \nn
f_7 &=& -\big(4\ V\ \big(4\ {\Phi'^4}\ V-2\ (78+5\ \Phi)
\ {\Phi'^2}\ \Phi''\ V \nn
&& +{{(18+\Phi)}^2}\ \Phi''\ V\ (\Phi''+24\ V) \nn
&& + 8\ (18+\Phi)\ {\Phi'^3}\ V'+2\ {{(18+\Phi)}^2}\ \Phi'
  \ (\Phi'''\ V-2\ (\Phi''+4\ V)\ V')\big)\big)\big/  \nn
&& \big(\big(2\ {\Phi'^2}-(24+\Phi)\ \Phi''\big)\ {{\big(-2
\ {\Phi'^2}+(18+\Phi)\ (\Phi''+8\ V)\big)}^2}\big) \nn
f_8 &=& \big(-56\ {\Phi'^4}\ V\ V'-4\ {{(18+\Phi)}^2}\ \Phi''
\ V\ (\Phi''+24\ V)\ V' \nn
&& - 4\ {\Phi'^2}\ V\ (3\ (18+\Phi)\ \Phi'''\ V \nn
&& -(246\ \Phi''+15\ \Phi\ \Phi''+288 \ V+16\ \Phi\ V)\ V') \nn
&& + 2\ {\Phi'^3}\ \big(9\ \Phi''\ {V^2}-8\ \big(6\ {V^3}+18
\ {V'^2}+\Phi\ {V'^2} \nn
&& +18\ V\ V''+\Phi\ V\ V''\big)\big)+\Phi'\   \nn
&& \big(9\ (10+\Phi)\ {\Phi''^2}\ {V^2} \nn
&& +8\ {{(18+\Phi)}^2}
\ V\ (-\Phi'''\ V'+8\ V\ V'')-8\ \Phi''\ \big(126\ {V^3}+3\ \Phi\ {V^3} \nn
&& - 324\ {V'^2}-36\ \Phi\ {V'^2}-{\Phi^2}\ {V'^2} \nn
&& -324\ V\ V''-36\ \Phi\ V\ V''-{\Phi^2}
\ V\ V''\big)\big)\big)\big/  \nn
&& \big(\big(2\ {\Phi'^2}-(24+\Phi)\ \Phi''\big)
  \ {{\big(-2\ {\Phi'^2}+(18+\Phi)\ (\Phi''+8\ V)\big)}^2}\big)
\ . \nonumber
\eea
Finally  substituting (\ref{vii4d}), (\ref{S1A}), (\ref{S2A})
and (\ref{S3A}) into the expression for the anomaly
(\ref{ano}), we obtain,
\bea
\label{AN1A}
T&=&-{1 \over 8\pi G}\left[
h_1 R^2 + h_2 R_{ij}R^{ij}
+ h_3 R^{ij}\partial_{i}\phi\partial_{j}\phi \right. \nn
&& + h_4 Rg^{ij}\partial_{i}\phi\partial_{j}\phi
+ h_5 R{1 \over \sqrt{-g}}\partial_{i}
(\sqrt{-g}g^{ij}\partial_{j}\phi) \nn
&& + h_6 (g^{ij}\partial_{i}\phi\partial_{j}\phi)^2
+ h_7 \left({1 \over \sqrt{-g}}\partial_{i}
(\sqrt{-g}g^{ij}\partial_{j}\phi)\right)^2 \nn
&& \left. + h_8 g^{kl}\partial_{k}\phi\partial_{l}\phi
{1 \over \sqrt{-g}}\partial_{i}(\sqrt{-g}g^{ij}\partial_{j}\phi)
\right] \\
h_1&=& \left[ 3\ \left\{(24-10\ \Phi)\ {\Phi'^6} \right. \right. \nn
&& + \big(62208+22464\ \Phi+2196\ {\Phi^2}+72
\ {\Phi^3}+{\Phi^4}\big)\ \Phi''\ {{(\Phi''+8\ V)}^2} \nn
&& + 2\ {\Phi'^4}\ \left\{\big(108+162\ \Phi+7\ {\Phi^2}\big)\
\Phi''+72\ \big(-8+14\ \Phi+{\Phi^2}\big)\ V\right\} \nn
&& - 2\ {\Phi'^2}\ \left\{\big(6912+2736\ \Phi+192
\ {\Phi^2}+{\Phi^3}\big)\ {\Phi''^2} \right. \nn
&& + 4\ \big(11232+6156\ \Phi+552\ {\Phi^2}
+13\ {\Phi^3}\big)\ \Phi''\ V \nn
&& \left. + 32\ \big(-2592+468\ \Phi+96\ {\Phi^2}+5
\ {\Phi^3}\big)\ {V^2}\right\} \nn
&& \left.\left. - 3\ (-24+\Phi)\ {{(6+\Phi)}^2}\ {\Phi'^3}\ (
\Phi'''+8\ V')\right\}\right] \big/  \nn
&& \left[16\ {{(6+\Phi)}^2}\ \left\{-2\ {\Phi'^2}
+(24+\Phi)\ \Phi''\right\}\ \left\{-2\ {\Phi'^2} \right.\right. \nn
&& \left.\left.+(18+\Phi)\ (\Phi''+8\ V)\right\}^2\right]\nn
h_2 &=&-\frac{3\ \left\{(12-5\ \Phi)\ {\Phi'^2}+(288+72\
\Phi+{\Phi^2})\ \Phi''\right\}}{8\ {{(6+\Phi)}^2}\
\left\{-2\ {\Phi'^2}+(24+\Phi)\ \Phi''\right\}} \nn
h_3 &=& -\big(3\ \big((12-5\ \Phi)\ {\Phi'^2}\ V+
\big(288+72\ \Phi+{\Phi^2}\big)\ \Phi''\ V \nn
&& +2\ \big(-144-18\ \Phi
+{\Phi^2}\big)\ \Phi'\ V'\big)\big)/  \nn
&& \big(4\ {{(6+\Phi)}^2}\ \big(-2\ {\Phi'^2}
+(24+\Phi)\ \Phi''\big)\big) \nn
h_4 &=& \big(-6\ (-12+5\ \Phi)\ {\Phi'^6}\ V \nn
&& +3 \ \big(62208+22464\ \Phi+2196
\ {\Phi^2}+72\ {\Phi^3}+{\Phi^4}\big)\
\Phi''\ V\ {{(\Phi''+8\ V)}^2} \nn
&& +2\ \big(-684-48\ \Phi+11
  \ {\Phi^2}\big)\ {\Phi'^5}\ V' \nn
&& + (6+\Phi)\ \Phi'\ \big(\big(-31104-2772\ \Phi+120\ {\Phi^2}+13
  \ {\Phi^3}\big)\ {\Phi''^2} \nn
&& + 8\ \big(-62208-7092\ \Phi-132\ {\Phi^2}+7
\ {\Phi^3}\big)\ \Phi''\ V \nn
&& + 384\ \big(-5184-504\ \Phi+6\ {\Phi^2}+{\Phi^3}
  \big)\ {V^2}\big)\ V' \nn
&& - (6+\Phi)\ {\Phi'^3}\ \big(9\ \big(-144-18\ \Phi+{\Phi^2}
  \big)\ \Phi'''\ V \nn
&& + \big(-3492\ \Phi''+252\ \Phi\ \Phi''+19\ {\Phi^2}\ \Phi'' \nn
&& -71712 \ V-4944\ \Phi\ V+208\ {\Phi^2}\ V\big)\ V'\big) \nn
&& + 6\ {\Phi'^4}\ \big(\big(108+162\ \Phi+7\ {\Phi^2}\big)\ \Phi''\ V+2
  \ \big(-288\ {V^2} \nn
&& + 504\ \Phi\ {V^2}+36\ {\Phi^2}\ {V^2}+864\ V''+252
\ \Phi\ V''+12\ {\Phi^2}\ V''-{\Phi^3}\ V''\big)\big) \nn
&& - 6\ {\Phi'^2}\ \big(\big(6912+2736\ \Phi+192
\ {\Phi^2}+{\Phi^3}\big)\ {\Phi''^2}\ V \nn
&& -82944\ {V^3}+14976\ \Phi\ {V^3} \nn
&& + 3072\ {\Phi^2}\ {V^3}+160\ {\Phi^3}\ {V^3}-15552
  \ \Phi'''\ V' \nn
&& -5400\ \Phi\ \Phi'''\ V'-468\ {\Phi^2}\ \Phi'''\ V' \nn
&& + 6\ {\Phi^3}\ \Phi'''\ V'+{\Phi^4}\ \Phi'''\ V'-124416
  \ {V'^2} \nn
&& -43200\ \Phi\ {V'^2}-3744\ {\Phi^2}\ {V'^2} \nn
&& + 48\ {\Phi^3}\ {V'^2}+8\ {\Phi^4}\ {V'^2}+124416\ V
  \ V'' \nn
&& +43200\ \Phi\ V\ V''+3744\ {\Phi^2}\ V\ V'' \nn
&& - 48\ {\Phi^3}\ V\ V''-8\ {\Phi^4}\ V\ V'' \nn
&& +\Phi''\ \big(44928\ {V^2}+24624
\ \Phi\ {V^2}+2208\ {\Phi^2}\ {V^2} \nn
&& + 52\ {\Phi^3}\ {V^2}+15552\ V''+5400\ \Phi\ V'' \nn
&& +468 \ {\Phi^2}\ V''-6\ {\Phi^3}\ V''
-{\Phi^4}\ V''\big)\big)\big)\big/  \nn
&& \big(8\ {{(6+\Phi)}^2}\ \big(-2\ {\Phi'^2}+(24+\Phi)\ \Phi''\big)
\ \big(-2\ {\Phi'^2} \nn
&& +(18+\Phi)\ (\Phi''+8\ V)\big)^2\big) \nn
h_5 &=& \big(\Phi'\ \big(-10\ {\Phi'^4}\ V+{\Phi'^2}\ V\ ((426+\Phi)
\ \Phi''-8\ (270+\Phi)\ V) \nn
&& + \Phi\ \Phi''\ V\ (-7\ (6+\Phi)\ \Phi''+8\ (174+5\ \Phi)\ V)
+ 12\ (-24+\Phi)\ {\Phi'^3}\ V' \nn
&& +6\ \big(-432-6
  \ \Phi+{\Phi^2}\big)\ \Phi'\ (\Phi'''\ V-\Phi''\ V')\big)\big)\big/  \nn &&
  \big(4\ \big(2\ {\Phi'^2}-(24+\Phi)\ \Phi''\big)
  \ {{\big(-2\ {\Phi'^2}+(18+\Phi)\ (\Phi''+8\ V)\big)}^2}\big) \nn
h_6 &=& \big(18\ (-12+5\ \Phi)\ {\Phi'^6}\ {V^2}+4
\ \big(2772+384\ \Phi-13\ {\Phi^2}\big)\ {\Phi'^5}\ V\ V' \nn
&& - \Phi''\ \big(3\ \big(124416+44928\ \Phi \nn
&& +4212\ {\Phi^2}+144\ {\Phi^3}
+{\Phi^4}\big)\ {\Phi''^2}\ {V^2} \nn
&& + 48\ \Phi''\ \big(124416\ {V^3}+44928\ \Phi
\ {V^3}+4212\ {\Phi^2}\ {V^3}+144\ {\Phi^3}\ {V^3}+{\Phi^4}\ {V^3} \nn
&& - 23328\ {V'^2}-10368\ \Phi\ {V'^2}-1584\ {\Phi^2}
\ {V'^2}-96\ {\Phi^3}\ {V'^2}-2\ {\Phi^4}\ {V'^2}\big) \nn
&& + 64\ V\ \big(373248\ {V^3}+134784
  \ \Phi\ {V^3} \nn
&& + 12636\ {\Phi^2}\ {V^3}+432\ {\Phi^3}\ {V^3}+3\ {\Phi^4}
  \ {V^3}-139968\ {V'^2} \nn
&& - 50544\ \Phi\ {V'^2}-4320\ {\Phi^2}\ {V'^2}+216
  \ {\Phi^3}\ {V'^2}+36\ {\Phi^4}\ {V'^2}+{\Phi^5}
  \ {V'^2}\big)\big) \nn
&& - (6+\Phi)\ {\Phi'^3}\ \big(9\ \big(-144-18
  \ \Phi+{\Phi^2}\big)\ \Phi'''\ {V^2} \nn
&& - 2\ V'\ \big(-17244\ \Phi''\ V-540\ \Phi\ \Phi''\ V+29
  \ {\Phi^2}\ \Phi''\ V-99360\ {V^2} \nn
&& + 1992\ \Phi\ {V^2}+212\ {\Phi^2}\ {V^2}+20736
  \ V''+3744\ \Phi\ V''-8\ {\Phi^3}\ V''\big)\big) \nn
&& + 2\ (6+\Phi)\ \Phi'\ V'\ \big(\big(62208+3708\ \Phi-24
  \ {\Phi^2}+{\Phi^3}\big)\ {\Phi''^2}\ V \nn
&& - 4\ \Phi''\ \big(-248832\ {V^2}-11736\ \Phi\ {V^2}+840
  \ {\Phi^2}\ {V^2}+34\ {\Phi^3}\ {V^2} \nn
&& + 46656\ V''+11016\ \Phi\ V''+468\ {\Phi^2}\ V''-18
  \ {\Phi^3}\ V''-{\Phi^4}\ V''\big) \nn
&& - 2\ \big(-432-6\ \Phi+{\Phi^2}\big)\
\big(4608\ {V^3}+192\ \Phi\ {V^3} \nn
&& +108\ \Phi'''\ V'+24\ \Phi\ \Phi'''\ V'
+{\Phi^2}\ \Phi'''\ V'+864\ {V'^2} \nn
&& + 192\ \Phi\ {V'^2}+8\ {\Phi^2}\ {V'^2}-1728\ V\ V'' \nn
&& -384\ \Phi\ V\ V''-16\ {\Phi^2}\ V\ V''\big)\big) \nn
&& - 2\ {\Phi'^4}\ \big(3\ \big(180+438\ \Phi+17
\ {\Phi^2}\big)\ \Phi''\ {V^2}+4\ \big(-4752\ {V^3} \nn
&& + 1116\ \Phi\ {V^3}+66\ {\Phi^2}\ {V^3}-3240
\ {V'^2}-1008\ \Phi\ {V'^2}-66\ {\Phi^2}\ {V'^2} \nn
&& + 2\ {\Phi^3}\ {V'^2}-2592\ V\ V''-756\ \Phi\ V\ V''-36
\ {\Phi^2}\ V\ V''+3\ {\Phi^3}\ V\ V''\big)\big) \nn
&& + 4\ {\Phi'^2}\ \big(6\ \big(2484+1197\ \Phi+84\ {\Phi^2}+2
  \ {\Phi^3}\big)\ {\Phi''^2}\ {V^2} \nn
&& + \Phi''\ \big(88128\ {V^3}+67608\ \Phi\ {V^3}+5040\ {\Phi^2}
\ {V^3}+90\ {\Phi^3}\ {V^3}-125712\ {V'^2} \nn
&& - 46656\ \Phi\ {V'^2}-4896\ {\Phi^2}\ {V'^2}-72\ {\Phi^3}
\ {V'^2}+5\ {\Phi^4}\ {V'^2}-46656\ V\ V'' \nn
&& - 16200\ \Phi\ V\ V''-1404\ {\Phi^2}\ V\ V''
+18\ {\Phi^3}\ V\ V''+3\ {\Phi^4}\ V\ V''\big) \nn
&& +  3\ V\ \big(-82944\ {V^3}+30528\ \Phi\ {V^3} \nn
&& +3840\ {\Phi^2}\ {V^3}+80\ {\Phi^3}\ {V^3}+15552\ \Phi'''\ V' \nn
&& + 5400\ \Phi\ \Phi'''\ V'+468\ {\Phi^2}\ \Phi'''\ V'
-6\ {\Phi^3}\ \Phi'''\ V'-{\Phi^4}
\ \Phi'''\ V'+72576\ {V'^2} \nn
&& + 28224\ \Phi\ {V'^2}+3360\ {\Phi^2}\ {V'^2}+112\ {\Phi^3}
  \ {V'^2}-124416\ V\ V'' \nn
&& - 43200\ \Phi\ V\ V''-3744\ {\Phi^2}\ V\ V''+48\ {\Phi^3}\ V
  \ V''+8\ {\Phi^4}\ V\ V''\big)\big)\big)\big/  \nn
&& \big(16\ {{(6+\Phi)}^2}\ \big(-2\ {\Phi'^2}+(24+\Phi)\ \Phi''\big)
\ \big(-2\ {\Phi'^2} \nn
&& +(18+\Phi)\ (\Phi''+8\ V)\big)^2\big) \nn
h_7 &=& -\big(V\ \big(84\ {\Phi'^4}\ V-8\ {{(18+\Phi)}^2}\ \Phi''\ V
\ (-3\ \Phi''+2\ (-12+\Phi)\ V) \nn
&& + {\Phi'^2}\ V\ \big(3\ \big(-876-40\ \Phi+{\Phi^2}\big)
\ \Phi''+8 \ {{(18+\Phi)}^2}\ V\big) \nn
&& - 4\ \big(-432-6\ \Phi+{\Phi^2}\big)\ {\Phi'^3}\ V' \nn
&& - (-24+\Phi)\ {{(18+\Phi)}^2}\ \Phi'\ (\Phi'''\ V-2
\ (\Phi''+4\ V)\ V')\big)\big)\big/  \nn
&& \big(\big(2\ {\Phi'^2}-(24+\Phi)\ \Phi''\big)\ {{\big(-2
\ {\Phi'^2}+(18+\Phi)\ (\Phi''+8\ V)\big)}^2}\big) \nn
h_8 &=& \big(-10\ {\Phi'^5}\ {V^2}
+4\ (-204+5\ \Phi)\ {\Phi'^4}\ V\ V'\nn
&& +32\ {{(18+\Phi)}^2}\ \Phi''\ V\ (-3\ \Phi''+2
\ (-12+\Phi)\ V)\ V' \nn
&& +2\ {\Phi'^2}\ V\ \big(3
  \ \big(-432-6\ \Phi+{\Phi^2}\big)\ \Phi'''\ V \nn
&& + \big(7416\ \Phi''+270\ \Phi\ \Phi''-11\
{\Phi^2}\ \Phi'' \nn
&& +1728\ V-480\ \Phi \ V-32\ {\Phi^2}\ V\big)\ V'\big) \nn
&& + {\Phi'^3}\ \big((426+\Phi)\ \Phi''\ {V^2}
 -8\ \big(270\ {V^3}+\Phi\ {V^3} \nn
&& + 432\ {V'^2}+6\ \Phi\ {V'^2}-{\Phi^2}\ {V'^2}+432
\ V\ V''+6\ \Phi\ V \ V''-{\Phi^2}\ V\ V''\big)\big) \nn
&& + \Phi'\ \big(-6\ \Phi\ \big(7\ {\Phi''^2}\ {V^2}
 -232\ \Phi''\ {V^3}+360\ \Phi'''\ V\ V' \nn
&& -360\ \Phi''\ {V'^2}-360\ \Phi''\ V\ V''-2880\ {V^2}\ V''\big) \nn
&& + 4\ {\Phi^3}\ \big(\Phi'''\ V\ V'-\Phi''\ {V'^2}-\Phi''\ V\ V''
 -8\ {V^2}\ V''\big) \nn
&& + 31104\ \big(-\Phi'''\ V\ V'+\Phi''\ {V'^2}
+\Phi''\ V\ V''+8\ {V^2}\ V''\big) \nn
&& - {\Phi^2}\ \big(7\ {\Phi''^2}\ {V^2}-40\ \Phi''
\ {V^3}-48\ \Phi'''\ V\ V'+48\ \Phi''\ {V'^2} \nn
&&+48\ \Phi''\ V\ V''+384\ {V^2}\ V''\big)\big)\big)\big/  \nn
&& \big(4\ \big(2\ {\Phi'^2}-(24+\Phi)\ \Phi''\big)
\ {{\big(-2\ {\Phi'^2}+(18+\Phi) \ (\Phi''+8\ V)\big)}^2}\big)
\ . \nonumber
\eea
The c functions proposed in this paper for $d=4$ case is given by
$h_1$ and $h_2$ by putting $\Phi'$ to vanish:
\bea
\label{CCbb}
c_1&=&{2\pi \over 3G}{62208+22464\Phi
+2196 \Phi^2 +72 \Phi^3+ \Phi^4 \over
(6+\Phi)^2(24+\Phi)(18+\Phi)} \nn
c_2&=&{3\pi \over G}{288+72 \Phi+ \Phi^2 \over
(6+\Phi)^2(24+\Phi)}
\eea
Note also that using of above condition on the zero value 
of dilatonic potential derivative on conformal boundary 
significally simplifies the conformal anomaly 
as many terms vanish.

\section{Comparison with other counterterm schemes}

In this Appendix, we compare the counter terms and 
the trace anomaly obtained here 
with those in ref.\cite{BGM} which appeared after this work has been 
submitted to hepth. For 
simplicity, we consider the case that the spacetime dimension is 4 and 
the boundary is flat and the metric $g_{ij}$ in (\ref{ib}) 
on the boundary is given by
\be
\label{AB1}
g_{ij}=F(\rho)\eta_{ij}\ .
\ee
We also assume the dilaton $\phi$ only depends on $\rho$: 
$\phi=\phi(\rho)$. This is exactly the case of ref.\cite{BGM}. 
Then the conformal anomaly (\ref{AN1}) vanishes on such background. 

Let us demonstrate that this is consistent with results of ref.\cite{BGM}.
In the metric (\ref{AB1}), the equation of motion 
(\ref{ii}) given by the variation over the dilaton $\phi$ 
and the Einstein equations in (\ref{iii}) have the 
following forms:
\bea
\label{AB2}
0&=&- {l \over 2\rho^3}F^2 \Phi'(\phi) - {2 \over l}\partial_\rho
\left({F^2 \over \rho}\partial_\rho\phi\right) \\
\label{AB3} 
0&=& {l^2 \over 12\rho^2}\left(\Phi(\phi) + {12 \over l^2}
\right) - {1 \over \rho^2} - {2 \over F}\partial_\rho^2 F 
+ {1 \over F^2}\left(\partial_\rho F\right)^2 
- {1 \over 2}\left(\partial_\rho \phi\right)^2 \\
\label{AB4}
0&=& {F \over 3\rho}\left(\Phi(\phi) + {12 \over l^2}
\right) - {2\rho \over l^2}\partial_\rho^2 F 
 - {2\rho \over l^2 F}\left(\partial_\rho F\right)^2 
+ {6 \over l^2}\partial_\rho F - {4F \over l^2 \rho}\ .
\eea
Eq.(\ref{AB3}) corresponds to $\mu=\nu=\rho$ component in 
(\ref{iii}) and (\ref{AB4}) to $\mu=\nu=i$. Other components 
equations in 
(\ref{iii}) vanish identically. 
Combining (\ref{AB3}) and (\ref{AB4}), we obtain
\bea
\label{AB5}
0&=& - {l^2 \over 4\rho^2}\left(\Phi(\phi) + {12 \over l^2}
\right) + {3 \over \rho^2} 
+ {3 \over F^2}\left(\partial_\rho F\right)^2 
 - {6 \over \rho F}\partial_\rho F
 - {1 \over 2}\left(\partial_\rho \phi\right)^2 \\
\label{AB6}
0&=& -{6 \over F}\partial_\rho^2 F 
+ {6 \over F^2}\left(\partial_\rho F\right)^2 
 - {6 \over \rho F}\partial_\rho F
 - 2\left(\partial_\rho \phi\right)^2 \ .
\eea
If we define a new variable $A$, which corresponds 
to the exponent in the warp factor  by 
\be
\label{AB7}
F=\rho\e^{2A}\ ,
\ee
Eq.(\ref{AB6}) can be rewritten as
\be
\label{AB8}
0=-{6 \over \rho}\partial_\rho \left(\rho \partial_\rho A\right)
 -\left(\partial_\rho \phi\right)^2\ .
\ee
Now we further define a new variable $B$ by 
\be
\label{AB9}
B\equiv \rho \partial_\rho A\ .
\ee
If ${\partial\phi \over \partial\rho}\neq 0$, 
we can regard $B$ as a 
function of $\phi$ instead of $\rho$ and one obtains 
\be
\label{AB10}
\partial_\rho B = {\partial B \over \partial\phi}
{\partial\phi \over \partial\rho}\ .
\ee
By substituting (\ref{AB9}) and (\ref{AB10}) into (\ref{AB8}), 
we find (by assuming ${\partial\phi \over \partial\rho}\neq 0$) 
\be
\label{AB11}
{\partial B \over \partial\phi} = - {1 \over 6\rho}
\partial_\rho \phi\ .
\ee
Using (\ref{AB5}) and (\ref{AB11}) (and also 
(\ref{AB7}) and (\ref{AB9})), we find that the dilaton 
equations motion (\ref{AB2}) is automatically satisfied. 

In \cite{BGM}, another counterterms scheme is proposed 
\be
\label{AB12}
S_{\rm BGM}^{(2)}={1 \over 16\pi G}\int d^4x 
\sqrt{-\hat g}\left\{{6u(\phi) \over l} + {l \over
2u(\phi)}R \right\}\ ,
\ee
instead of (\ref{IIIb}). Here $u$ is obtained in terms of this 
paper as follows:
\be
\label{AB13}
u(\phi)^2=1 + {l^2 \over 12}\Phi(\phi)\ .
\ee
Then based on the counter terms in (\ref{AB12}), the following 
expression of the trace anomaly is given in \cite{BGM}:\footnote{
The radial coordinate $r$ in \cite{BGM} is related to $\rho$ 
by $dr = {ld\rho \over 4\rho}$. Therefore $\partial_r = - {2\rho 
\over l}\partial_\rho$, especially $\partial_r A = -{2\rho 
\over l}\partial_\rho A = - {2 \over l}B$. }
\be
\label{AB14}
T={3 \over 2\pi G l}(-2 B - u)\ .
\ee
The above trace anomaly was evaluated for fixed but finite $\rho$. 
If the boundary is asymptotically AdS, $F$ in (\ref{AB1}) 
goes to a constant $F\rightarrow F_0$ ($F_0$: a constant). 
Then from (\ref{AB7}) and (\ref{AB9}, we find the behaviors of 
$A$ and $B$ as 
\be
\label{AB15}
A\rightarrow {1 \over 2}\ln {F_0 \over \rho}\ ,
\quad B\rightarrow -{1 \over 2}\ .
\ee
Then (\ref{AB11}) tells that the dilaton $\phi$ becomes a 
constant. Then (\ref{AB5}) tells that 
\be
\label{AB16}
u=\sqrt{1 + {l^2 \over 12}\Phi(\phi)}\rightarrow 1\ .
\ee
Eqs.(\ref{AB15}) and (\ref{AB16}) tell that the trace anomaly 
(\ref{AB14}) vanishes on the boundary.  Thus, we demonstrated that trace 
anomaly of \cite{BGM} vanishes in the UV limit what is expected 
also from AdS/CFT correspondence.

We should note that the trace anomaly (\ref{AN1}) is evaluated 
on the boundary, i.e., in the UV limit.  
We evaluated the anomaly by expandind the action in 
the power series of $\epsilon$ in (\ref{vibc}) and subtracting 
the divergent terms in the limit of $\epsilon\rightarrow 0$. 
If we evaluate the anomaly for finite $\rho$ as in \cite{BGM}, 
the terms with positive power of $\epsilon$ in the expansion do 
not vanish and we would obtain non-vanishing trace anomaly in 
general. Thus, the trace anomaly obtained in this paper does not not 
have any contradiction with that in \cite{BGM}.

\section{Remarks on boundary values}

 From the leading order term in the equations of motion
\be
\label{eqm1}
0=-\sqrt{-\hat{G}}{\partial \Phi(\phi_{1},\cdots ,\phi_{N})
 \over \partial \phi_{\beta}} - \partial_{\mu }\left(\sqrt{-\hat{G}}
\hat{G}^{\mu \nu}\partial_{\nu} \phi_{\beta} \right)\ ,
\ee
which are given by variation of the action (\ref{mul})
\bea
\label{mul}
S={1 \over 16\pi G}\int_{M_{d+1}} d^{d+1}x \sqrt{-\hat G}
\left\{ \hat R - \sum_{\alpha=1 }^{N} {1 \over 2 }
(\hat\nabla\phi_{\alpha})^2
+\Phi (\phi_{1},\cdots ,\phi_{N} )+4 \lambda ^2 \right\}.&&
\eea
with respect to $\phi_{\alpha}$, we obtain
\be
\label{fco}
{\partial \Phi(\phi_{(0)}) \over \partial \phi_{(0)\alpha} } =0 .
\ee
The equation (\ref{fco}) gives one of the necessary conditions 
that the spacetime is asymptotically AdS. The equation 
(\ref{fco}) also looks like a constraint that the 
boundary value $\phi_{(0)}$ must take a special value 
satisfying (\ref{fco}) for the general fluctuations but it 
is not always correct. The condition $\phi=\phi_{(0)}$ at 
the boundary is, of course, the boundary condition, which 
is not a part of the equations of motion. Due to the boundary 
condition, not all degrees of freedom of $\phi$ are 
dynamical. Here the boundary value $\phi_{(0)}$ is, of 
course, not dynamical. This tells that we should not impose the 
equations given only by the variation over $\phi_{(0)}$. The 
equation (\ref{fco}) is, in fact, only given by the variation 
of $\phi_{(0)}$. In order to understand the situation, we choose 
the metric in the following form
\be
\label{ibB}
ds^2\equiv\hat G_{\mu\nu}dx^\mu dx^\nu
= {l^2 \over 4}\rho^{-2}d\rho d\rho + \sum_{i=1}^d
\hat g_{ij}dx^i dx^j \ ,\quad
\hat g_{ij}=\rho^{-1}g_{ij}\ ,
\ee
(If $g_{ij}=\eta_{ij}$, the boundary of AdS lies at $\rho=0$.) 
and we use the regularization for the action (\ref{mul})
by introducing the infrared cutoff
$\epsilon$ and replacing
\be
\label{vibcB}
\int d^{d+1}x\rightarrow \int d^dx\int_\epsilon d\rho \ ,\ \
\int_{M_d} d^d x\Bigl(\cdots\Bigr)\rightarrow
\int d^d x\left.\Bigl(\cdots\Bigr)\right|_{\rho=\epsilon}\ .
\ee
Then the action (\ref{mul}) has the following form:
\be
\label{mulb}
S={l \over 16\pi G }{1 \over d}\epsilon^{-{d \over 2}}
\int_{M_d}d^d x \sqrt{-\hat g_{(0)}}
\left\{ 
\Phi (\phi_{1(0)},\cdots ,\phi_{N(0)} ) -{8 \over l^2} \right\}
+ {\cal O}\left(\epsilon^{-{d \over 2}+1}\right)\ .
\ee
Then it is clear that Eq.(\ref{fco}) can be derived only from 
the variation over $\phi_{(0)}$ but not other components 
$\phi_{(i)}$ ($i=1,2,3,\cdots$). Furthermore, if we add the 
surface counterterm $S_b^{(1)}$ 
\be
\label{Sb1} 
S_b^{(1)}=-{1 \over 16\pi G}{d \over 2}\epsilon^{-{d \over 2}}
\int_{M_d}d^d x \sqrt{-\hat g_{(0)}}
\Phi (\phi_{1(0)},\cdots ,\phi_{N(0)} )
\ee
to the action (\ref{mul}), the first $\phi_{(0)}$ dependent 
term in (\ref{mulb}) is cancelled and we find that Eq.(\ref{fco}) 
cannot be derived from the variational principle. The surface 
counterterm in (\ref{Sb1}) is a part of the surface counterterms, 
which are necessary to obtain the well-defined AdS/CFT 
correspondence. Since the volume of AdS is infinte, the action 
(\ref{mul}) contains divergences, a part of which appears in 
(\ref{mulb}). Then in order that we obtain the well-defined 
AdS/CFT set-up, we need the surface counterterms to cancell the 
divergence.

\end{document}